\documentclass[final,3p,times]{elsarticle}

\usepackage{amssymb}
\usepackage{amsmath}

\journal{Computer Methods in Applied Mechanics and Engineering}

\usepackage{tikz}
\usetikzlibrary{arrows.meta}
\usepackage{subfig}
\newcommand{\mbf}{\mathbf}  
\def\Tr{^\mathsf{T}} 
\newcommand{\tss}{\textsuperscript}  
\usepackage{algorithm}
\usepackage{algorithmic} 
\usepackage{booktabs}

\begin{document}

\begin{frontmatter}



\title{Space-Time Multigrid Methods Suitable for Topology Optimisation of Transient Heat Conduction}


\author[sduLabel]{Magnus Appel\corref{cor1}} 
\ead{magap@sdu.dk}
\author[sduLabel]{Joe Alexandersen} 
\ead{joal@sdu.dk}

\cortext[cor1]{Corresponding author}

\affiliation[sduLabel]{organization={University of Southern Denmark},
            addressline={Campusvej 55}, 
            city={Odense},
            postcode={5230}, 
            country={Denmark}}

\begin{abstract}

This paper presents Space-Time MultiGrid (STMG) methods which are suitable for performing topology optimisation of transient heat conduction problems. The proposed methods use a pointwise smoother and uniform Cartesian space-time meshes. For problems with high contrast in the diffusivity, it was found that it is beneficial to define a coarsening strategy based on the geometric mean of the minimum and maximum diffusivity. However, other coarsening strategies may be better for other smoothers. Several methods of discretising the coarse levels were tested. Of these, it was best to use a method which averages the thermal resistivities on the finer levels. However, this was likely a consequence of the fact that only one spatial dimension was considered for the test problems. A second coarsening strategy was proposed which ensures spatial resolution on the coarse grids. Mixed results were found for this strategy. The proposed STMG methods were used as a solver for a one-dimensional topology optimisation problem. In this context, the adjoint problem was also solved using the STMG methods. The STMG methods were sufficiently robust for this application, since they converged during every optimisation cycle. It was found that the STMG methods also work for the adjoint problem when the prolongation operator only sends information forwards in time, even although the direction of time for the adjoint problem is backwards. 

\end{abstract}


\begin{highlights}
\item Space-time multigrid methods for transient heat conduction.
\item Semi-coarsening strategies suitable for high contrast problems. 
\item Different methods of interpolating the temperature and methods of discretising the coarse levels. 
\item Demonstration of robustness by applying the methods to a one-dimensional topology optimisation problem. 
\end{highlights}


\end{frontmatter}



\section{Introduction}

\subsection{Motivation}

Topology optimisation methods are computational design methods which aim to find optimal material distributions for various engineering applications \cite{bendsoe2003_book, deaton2014_survey}. One such application is optimising transient heat conduction \cite{Zhuang2013_TOot, Zhuang2014_globalcompliance, Dbouk2017_review,Wu2021_TTDE}, which will be the focus of this article. Transient effects have previously been demonstrated to have an impact on optimised designs \cite{wu2019_minmaxTemp,Zeng2020} and are important in practise since many physical situations are transient by nature, such as the workload and heat generation of electronics chips \cite{Zeng2020}.

Topology optimisation methods can be hindered by their large computational costs, caused by the fact that the system being optimised has to be simulated multiple times during the iterative optimisation process \cite{bendsoe2003_book}. This especially becomes an issue for time-dependent problems, where the computational time of each simulation can easily reach orders of hours or days for realistic problems, leading to optimisation times of e.g. 12 days for two-dimensional natural convection \cite{Coffin2016} and ``several days'' for three-dimensional photonics \cite{Elesin2012}.
As such, there is a demand for fast and efficient solver algorithms.

\subsection{Literature}

A wide variety of methods have already been applied for the purpose of accelerating topology optimisation \cite{mukherjee2021_accelerating}, although mainly for steady problems. This includes multigrid methods, such as Geometric MultiGrid (GMG) \cite{amir2014_topOptMG, aage2015_openSourceTopOpt} and Algebraic MultiGrid (AMG) methods \cite{HerreroPerez2021_gpuAMG, Peetz2021_multigridCompare}. These multigrid methods have been used as preconditioners for outer solver algorithms, such as the conjugate gradient method \cite{Hestenes1952_CG} and GMRES \cite{Saad1986_GMRES}. These methods have the benefit of being scalable and parallelisable. However, the performance of multigrid methods are generally degraded if the considered problem contains small features and high contrasts in the coefficients \cite{amir2014_topOptMG}, which is often the case when performing topology optimisation. Despite this, multigrid methods have been used to solve a variety of large-scale topology optimisation problems \cite{Alexandersen2016_3dnatconv, Aage2017_gigascale, Hoghoj2020_hex, Traff2021_shell, Herrero-Perez2023_adaptiveGMG, Luo2024_isogeoGMG, zhou2025robustsolverlargescaleheat}. 
Some work has also been done to develop contrast-independent multilevel preconditioners for topology optimisation \cite{Lazarov2014_multiscale,Alexandersen2015_multiscale,Alexandersen2015_multiscale2,Zambrano2021_multiscale}.

Most of the work on parallelising topology optimisation methods has focused on parallelising the computation with respect to space, where each processor works on one patch of the spatial domain. 
Although applicable to transient problems \cite{Kristiansen2022}, this method of parallelising the computation has a limit to the amount of speed-up that it can provide, since the time-stepping method is still sequential. To address this problem, parallel-in-time methods can be used. 

A number of different parallel-in-time methods have been devised and studied as covered by the review papers by \citet{Gander2015_review} and \citet{Ong2020_review}. A few examples include Parareal \cite{parareal_orig}, multigrid reduction in time \cite{mgrit_original}, the parallel time-stepping method \cite{womble1990_parTimeStep}, and PFASST \cite{pfasst_original}. Some of these methods have been applied to optimisation problems, like size optimisation \cite{hahne2023_pintMotorOpt} and optimal control problems \cite{du2013_pararealPrecond, Minion2018_pfasstOptCtrl, Gunther2019_mgritOptCtrl}. However, to the authors' knowledge, only two parallel-in-time methods have been applied to topology optimisation in the literature. Firstly, there is the parallel local-in-time method \cite{Theulings_PLT}, which has the disadvantage that it becomes unreliable when using too many processors. Secondly, there is the one-shot Parareal method of topology optimisation \cite{appel2024_oneShotPar}, which has the disadvantage that it has poor scalability. To address these problems, this article focuses on Space-Time MultiGrid (STMG) methods \cite{hackbusch1983_stmg}, because these have been reported to be among the fastest parallel-in-time methods \cite{falgout2017_stmgComparisons}. These methods work similarly to spatial multigrid methods, in the sense that they define a hierarchy of coarse grids, after which they use smoothing schemes and coarse grid corrections to solve the considered problem. However, STMG methods are applied to space-time domains, where the time axis is treated as being another dimension of the domain, as opposed to purely spatial domains. 

Space-time domains are inherently anisotropic, since the dimension of time is qualitatively different from the dimensions of space. This means that STMG methods are subject to challenges which appear in the context of spatial multigrid methods for anisotropic problems. For example, if standard coarsening (where all dimensions of the problem are coarsened simultaneously) and pointwise smoothers are used, then this can lead to very slow convergence rates \cite{horton1995_stMultigrid}. This can be avoided by using semi-coarsening in the direction of strongest coupling \cite{horton1995_stMultigrid} or choosing different smoothers depending on the direction of strongest coupling \cite{franco2018_stmg_with_CN}. In the context of STMG methods for diffusion problems, the direction of strongest coupling is determined by a parameter, $\lambda$, which is defined as $\lambda = D \Delta t/ \Delta x^2$, where $D$ is the diffusivity coefficient, $\Delta t$ is the length of the time steps, and $\Delta x$ is the length of the spatial steps.  If $\lambda$ is small, then the coupling is strongest in the temporal direction, while if $\lambda$ is large, then the coupling is strongest in the spatial directions. This parameter, $\lambda$, will be referred to as the ``anisotropy parameter" in this article, but it has also been referred to as the ``discrete anisotropy" \cite{horton1995_stMultigrid} and the ``discretisation parameter" \cite{Gander2016_stmg_br}.

For realistic heat conduction problems involving both thermal conductors and insulators, the diffusivity becomes heterogeneous, and the contrast in the diffusivity can easily be on the order of $10^3$, meaning that the contrast in $\lambda$ can also be on the order of $10^3$. Therefore, the direction of strongest coupling becomes different in different parts of the domain. As such, it is less clear how to define good semi-coarsening strategies or smoothing schemes for these heterogeneous problems. 

The authors only know of four works which consider STMG methods applied to heterogeneous problems. \citet{Li2022_stmgForPorousFlow} applied an STMG method to high-contrast, non-linear, two-phase flow in a porous medium. For this, adaptive refinement of the space-time mesh was used instead of adaptive coarsening. \citet{margenberg2024_stmgHetWave} applied an STMG method to wave propagation problems where the propagation speed was variable over space. A pointwise smoother and semi-coarsening strategy was used, but the coarsening strategy was independent of the material parameters. \citet{schwalsberger2018_eddyThesis} and \citet{neumuller2019_stmgEddy} applied an STMG method to high-contrast eddy current problems, which are diffusion problems. For this, coarsening strategies were used where the coarsening type is decided based on quantities which are analogous to the minimum value of $\lambda$.

\subsection{Contributions}


This paper proposes space-time multigrid methods which are suitable for performing topology optimisation of transient heat conduction problems. In particular, it proposes coarsening strategies which are suitable for high-contrast problems, investigates different methods of defining the discretisation on the coarse grids, and demonstrates the robustness of these strategies and methods.

\subsection{Paper layout}

The paper is structured as follows: 
Section \ref{sec: gov eq} presents the considered governing equations and discretisation; 
Section \ref{sec: stmg def} presents the STMG method used to solve the discretised equations, and explains why the anisotropy parameter, $\lambda$, is important;
Section \ref{sec: eff lambda} presents an ``effective anisotropy parameter" which works well for defining a semi-coarsening strategy for high-contrast problems;
Section \ref{sec: different redisc methods} presents different interpolation methods and methods of re-discretising the coarse grids of the STMG method;
Section \ref{sec: new coar strat} presents a semi-coarsening strategy which ensures spatial resolution on the coarse grids; 
Section \ref{sec: opt prob application} applies the STMG method to an optimisation problem that is similar to a topology optimisation problem; 
and Section \ref{sec: conclusion} gives a conclusion.

\section{Governing equations}
\label{sec: gov eq}

Topology optimisation problems involving heat conduction usually consider two or three dimensions of space. However, for the sake of simplicity, this work considers transient heat conduction in one spatial dimension governed by the following partial differential equation (PDE) and boundary conditions: 
\begin{subequations}
\begin{align}
    c \frac{\partial T}{\partial t} - \frac{\partial}{\partial x} \left( k \frac{\partial T}{\partial x} \right) &= q
    \quad \text{for } 0 \leq x \leq L, 0 \leq t \leq t_T
    \\
    T &= 0 \quad \text{at } t=0
    \\
    T &= 0 \quad \text{at } x=0
    \\
    \frac{\partial T}{\partial x} &= 0 \quad \text{at } x=L
\end{align}
\end{subequations}
where $T=T(x,t)$ is the temperature field, $c=c(x)$ is the volumetric heat capacity, $k=k(x)$ is the thermal conductivity, $q=q(x,t)$ is the imposed external heat load per unit volume, $L$ is the length of the domain, and $t_T$ is the terminal time of the problem. These one-dimensional heat conduction problems could, for example, represent heat flow through a long rod or a planar wall.

\subsection{Parametrisation of the material properties}
\label{sec: simp scheme}

This work focuses on density-based topology optimisation where the domain is filled with two materials: 
\begin{enumerate}
    \item An insulating material where $k=k_\mathrm{ins}$ and $c=c_\mathrm{ins}$;
    \item A conducting material where $k=k_\mathrm{con}$ and $c=c_\mathrm{con}$.
\end{enumerate}
The placement of these materials is parametrised by a function, $\chi(x)$, which is referred to as ``the design field". Inside the insulator, $\chi(x)=0$, and inside the conductor, $\chi(x)=1$. In density-based topology optimisation, the design field is relaxed to be continuous and allowed to vary between 0 and 1. As such, there are some points where $0 < \chi(x) < 1$, which represents a mix of the insulating and conducting material. At every point in the domain, the conductivity and heat capacity are defined according to a Solid Isotropic Material with Penalisation (SIMP) scheme \cite{bendsoe1989_simp, rozvany1992_simp} of the following form:
\begin{subequations}
\begin{align}
    k(x) &= k_\mathrm{SIMP}(\chi(x))
    \\ 
    c(x) &= c_\mathrm{SIMP}(\chi(x))
\end{align}
\end{subequations}
where the functions $k_\mathrm{SIMP}$ and $c_\mathrm{SIMP}$ are defined as follows:
\begin{subequations} 
\begin{align}
    \label{eq: SIMP k}
    k_\mathrm{SIMP}(\chi) = k_\mathrm{ins} + (k_\mathrm{con} - k_\mathrm{ins}) \chi^{p_k}
    \\ 
    \label{eq: SIMP c}
    c_\mathrm{SIMP}(\chi) = c_\mathrm{ins} + (c_\mathrm{con} - c_\mathrm{ins}) \chi^{p_c}
\end{align}
\end{subequations}
where $p_k$ and $p_c$ are penalty powers assigned to the conductivity and heat capacity, respectively. This paper exclusively considers $p_k=3$ and $p_c=2$, because these values have been demonstrated to work well for minimising areas where $0 < \chi(x) < 1$ when performing topology optimisation of transient heat flow in two spatial dimensions \cite{wu2019_minmaxTemp}.

\subsection{Discretisation}

The problem is discretised in space using a Galerkin finite element method with equally sized elements. The number of elements is denoted $N_\mathrm{el}$, so the width of each element is $\Delta x \equiv L/N_\mathrm{el}$, and the $x$-coordinate of the centre of the $e$\tss{th} element is $x_e \equiv \left( e + 1/2 \right)\Delta x $. Linear shape functions are used for $T$ and element-wise constant functions are used for $q$, $\chi$, $c$, and $k$. The problem is discretised in time using the backward Euler method, and the solution is evaluated at time points $t_0, t_1, \hdots t_{N_t}$, where $N_t$ is the number of time steps. These time points are evenly spaced, so $\Delta t \equiv t_T/N_t$ and $t_n \equiv n\cdot \Delta t$. As such, the discretised problem is of the following form:
\begin{subequations}
\begin{align}
    \mbf{T}_0 &= \mbf{0}
    \\
    \mbf{C} \frac{\mbf{T}_n - \mbf{T}_{n-1}}{\Delta t} + \mbf{K} \mbf{T}_n &= \mbf{q}_n 
    \quad
    \text{for } n = 1, 2, \hdots N_t-1, N_t
\end{align}
\end{subequations}
where $\mbf{T}_n$ is the vector containing the degrees of freedom of the temperature at $t=t_n$, $\mbf{C}$ is the spatial heat capacity matrix, $\mbf{K}$ is the spatial thermal conductivity matrix, and $\mbf{q}_n$ is the external heat load vector at $t=t_n$. The discretized problem can be written in all-at-once matrix form like so:
\begin{align}
    \begin{pmatrix}
        W_\mathrm{Diri} \cdot \mbf{I} & \mbf{0} & \mbf{0} & \hdots & \mbf{0} & \mbf{0} \\
        -\mbf{C}/\Delta t  & \mbf{C}/\Delta t + \mbf{K} & \mbf{0} & \hdots & \mbf{0} & \mbf{0} \\
        \mbf{0} & -\mbf{C}/\Delta t  & \mbf{C}/\Delta t + \mbf{K} & \hdots & \mbf{0} & \mbf{0} \\
        \vdots & \vdots & \vdots & \ddots & \vdots & \vdots \\
        \mbf{0} & \mbf{0} & \mbf{0} & \hdots & \mbf{C}/\Delta t + \mbf{K} & \mbf{0} \\
        \mbf{0} & \mbf{0} & \mbf{0} & \hdots & -\mbf{C}/\Delta t & \mbf{C}/\Delta t + \mbf{K}
    \end{pmatrix}
    \begin{pmatrix}
        \mbf{T}_0 \\ \mbf{T}_1 \\ \mbf{T}_2 \\ \vdots \\ \mbf{T}_{N_t-1} \\ \mbf{T}_{N_t}
    \end{pmatrix}
    &=
    \begin{pmatrix}
        \mbf{0} \\ \mbf{q}_1 \\ \mbf{q}_2 \\ \vdots \\ \mbf{q}_{N_t-1} \\ \mbf{q}_{N_t}
    \end{pmatrix}
\end{align}
where $\mbf{I}$ is the identity matrix and $W_\mathrm{Diri}$ is a scalar weight assigned to the residual associated with the Dirichlet boundary conditions. For more information about the value of $W_\mathrm{Diri}$ and the treatment of Dirichlet boundary conditions, see \ref{app: handling diri bc}. The above system of equations is shortened down to the following:
\begin{align}
    \label{eq: all-at-once matrix form short}
    \mbf{J} \mbf{u} &= \mbf{b}
\end{align}
where $\mbf{u} = \left( \mbf{T}_0\Tr \,\, \mbf{T}_1\Tr \,\, \mbf{T}_2\Tr \,\, \hdots \,\, \mbf{T}_{N_t}\Tr \right) \Tr$ is the vector of unknowns, $\mbf{b} = \left( \mbf{0} \,\, \mbf{q}_1\Tr \,\, \mbf{q}_2\Tr \,\, \hdots \,\, \mbf{q}_{N_t}\Tr \right) \Tr$ is the source term vector, and $\mbf{J}$ is the all-at-once system matrix. This all-at-once space-time discretisation is referred to as the Backward Euler Finite Element (BE-FE) discretisation. If $k$ and $c$ are constant over space, then the space-time stencil associated with $\mbf{J}$ is the following at the interior points of the space-time domain:
\begin{align}
    \label{eq: stencil of J 1}
    c \frac{\Delta x}{\Delta t} \cdot
    \begin{bmatrix}
        0 & 0 & 0 \\ 
        1/6 & 2/3 & 1/6 \\
        -1/6 & -2/3 & -1/6
    \end{bmatrix}
    +
    k \frac{1}{\Delta x} \cdot
    \begin{bmatrix}
        0 & 0 & 0 \\ 
        -1 & 2 & -1 \\
        0 & 0 & 0
    \end{bmatrix}
\end{align}
In this stencil, the vertical direction represents the time direction and the horizontal direction represents the spatial dimension. The top row of this stencil is 0 because the backward Euler method does not transfer information backwards in time.

\section{Space-time multigrid method} 
\label{sec: stmg def}

In this work, STMG methods are used to solve Equation (\ref{eq: all-at-once matrix form short}). This section describes the STMG method that is studied in Section \ref{sec: stmg def} and \ref{sec: eff lambda} of this paper. In Section \ref{sec: different redisc methods}, it will be modified by introducing different interpolation methods and matrix reassembly methods.

\subsection{Multigrid method components}
\label{sec: stmg components def}

The damped pointwise Jacobi method is used as the smoother. One step of this smoother is applied on the $l$\tss{th} level of the multigrid method by performing the following assignment: 
\begin{align}
    \mbf{u}^{(l)} \gets \mbf{u}^{(l)} + \omega \cdot \left( \mbf{D}^{(l)} \right)^{-1} \left( \mbf{b}^{(l)} - \mbf{J}^{(l)} \mbf{u}^{(l)} \right)
\end{align}
where $\omega$ is the damping factor, and $\mbf{D}$ is the diagonal matrix containing the main diagonal of $\mbf{J}$. Also, the superscripts $^{(l)}$ indicate that these variables pertain to the $l$\tss{th} level of the multigrid method. This paper exclusively considers $\omega = 1/2$ because this value was observed to work reasonably well in practice. 

\begin{figure}
\centering
\subfloat[Semi-coarsening in space ($x$-coarsening).]{
\begin{tikzpicture}[scale=0.4]
\foreach \i in {0,...,6}  \draw[thin, solid, black] (\i, 0) -- (\i, 6); 
\foreach \i in {0,...,6}  \draw[thin, solid, black] (0, \i) -- (6, \i); 
\draw[black, thin, {}-{Stealth[scale=1.3]} ] (0,-1)  -- (6,-1) node[anchor=west] {$x$}; 
\draw[black, thin, {}-{Stealth[scale=1.3]} ] (-1,0)  -- (-1,6) node[anchor=south] {$t$}; 
\draw[black, very thick, {}-{Latex[scale=1]} ] (7,3)  -- (9,3); 
\foreach \i in {0,...,3}  \draw[thin, solid, black] (2*\i + 10, 0) -- (2*\i + 10, 6); 
\foreach \i in {0,...,6}  \draw[thin, solid, black] (0 + 10, \i) -- (6 + 10, \i); 
\end{tikzpicture}
} \hspace{7mm}
\subfloat[Semi-coarsening in time ($t$-coarsening).]{
\begin{tikzpicture}[scale=0.4]
\foreach \i in {0,...,6}  \draw[thin, solid, black] (\i, 0) -- (\i, 6); 
\foreach \i in {0,...,6}  \draw[thin, solid, black] (0, \i) -- (6, \i); 
\draw[black, thin, {}-{Stealth[scale=1.3]} ] (0,-1)  -- (6,-1) node[anchor=west] {$x$}; 
\draw[black, thin, {}-{Stealth[scale=1.3]} ] (-1,0)  -- (-1,6) node[anchor=south] {$t$}; 
\draw[black, very thick, {}-{Latex[scale=1]} ] (7,3)  -- (9,3); 
\foreach \i in {0,...,6}  \draw[thin, solid, black] (\i + 10, 0) -- (\i + 10, 6); 
\foreach \i in {0,...,3}  \draw[thin, solid, black] (0 + 10, 2*\i) -- (6 + 10, 2*\i); 
\end{tikzpicture}
}

\subfloat[Full space-time coarsening.]{
\begin{tikzpicture}[scale=0.4]
\foreach \i in {0,...,6}  \draw[thin, solid, black] (\i, 0) -- (\i, 6); 
\foreach \i in {0,...,6}  \draw[thin, solid, black] (0, \i) -- (6, \i); 
\draw[black, thin, {}-{Stealth[scale=1.3]} ] (0,-1)  -- (6,-1) node[anchor=west] {$x$}; 
\draw[black, thin, {}-{Stealth[scale=1.3]} ] (-1,0)  -- (-1,6) node[anchor=south] {$t$}; 
\draw[black, very thick, {}-{Latex[scale=1]} ] (7,3)  -- (9,3); 
\foreach \i in {0,...,3}  \draw[thin, solid, black] (2*\i + 10, 0) -- (2*\i + 10, 6); 
\foreach \i in {0,...,3}  \draw[thin, solid, black] (0 + 10, 2*\i) -- (6 + 10, 2*\i); 
\end{tikzpicture}
}
\caption{Illustrations of the three different types of coarsening which are defined in Section \ref{sec: stmg components def}. }
\label{fig: coar types sketches}
\end{figure}
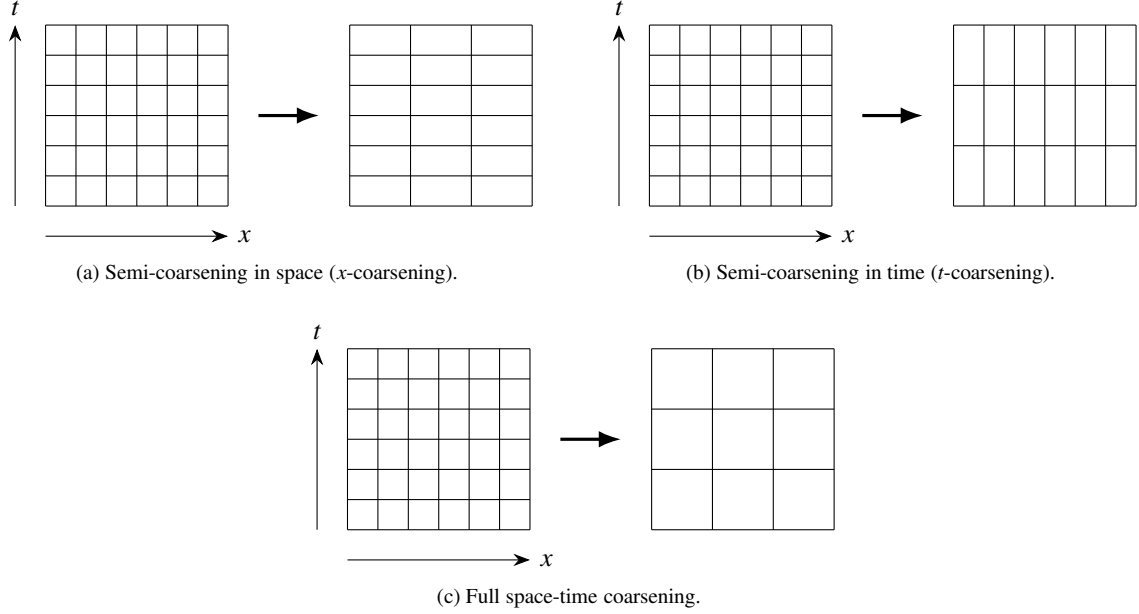

Three different types of coarsening are considered: 
\begin{itemize}
    \item Semi-coarsening in space ($x$-coarsening), where $\Delta x$ is doubled on the next coarse level while $\Delta t$ is unchanged;
    \item Semi-coarsening in time ($t$-coarsening), where $\Delta t$ is doubled on the next coarse level while $\Delta x$ is unchanged;
    \item Full space-time coarsening, where both $\Delta t$ and $\Delta x$ are doubled. 
\end{itemize}
These coarsening types are illustrated in Figure \ref{fig: coar types sketches}. The stencils for the prolongation operators associated with these coarsening types are:
\begin{align}
    x\text{-coarsening: } 
    \frac{1}{2} 
    \left] \begin{matrix}
        0 & 0 & 0 \\
        1 & 2 & 1 \\
        0 & 0 & 0
    \end{matrix} \right[ 
    \quad
    t\text{-coarsening: } 
    \left] \begin{matrix}
        0 & 1 & 0 \\
        0 & 1 & 0 \\
        0 & 0 & 0
    \end{matrix} \right[ 
    \quad
    \text{Full coarsening: } 
    \frac{1}{2} 
    \left] \begin{matrix}
        1 & 2 & 1 \\
        1 & 2 & 1 \\
        0 & 0 & 0
    \end{matrix} \right[ 
\end{align}
These stencils follow the same convention for the directions as the stencil in Equation (\ref{eq: stencil of J 1}). However, by convention, the above stencils indicate where the information is distributed to, whereas the stencil in Equation (\ref{eq: stencil of J 1}) indicates where information is taken from. As such, these prolongation operators are causal, since they do not transfer any information backwards in time. The restriction operators are of the following form:
\begin{align}
    \mbf{R}^{(l)} &= s^{(l)} \cdot { \mbf{P}^{(l)} }\Tr 
\end{align}
where $\mbf{R}^{(l)}$ is the restriction operator which transfers information from the $l$\tss{th} level to the $(l+1)$\tss{th} level, $\mbf{P}^{(l)}$ is the prolongation operator which transfers information in the opposite direction, and $s^{(l)}$ is a scalar which is used to ensure that the coarse grid correction is consistent with the PDE. For $x$-coarsening, $s^{(l)}=1$, while for full coarsening and $t$-coarsening, $s^{(l)}=1/2$. These scalars account for two effects: Firstly, the restriction operator is scaled differently from the prolongation operator, because the restriction operator acts as a weighted average while the prolongation operator acts to copy information. Secondly, the residual associated with the BE-FE discretisation is weighted differently between different levels of the multigrid method.

The system matrices on the coarse levels are reassembled according to the BE-FE discretisation method, rather than using Galerkin projection which is commonly used in spatial multigrid methods for topology optimisation \cite{amir2014_topOptMG, aage2015_openSourceTopOpt}. When applying $x$-coarsening or full space-time coarsening, the coefficients $k$ and $c$ on the coarse levels are computed as being averages of $k$ and $c$ on the finer levels like so:
\begin{subequations}
    \begin{align}
        k_e^{(l+1)} &= \frac{ k_{2e}^{(l)} + k_{2e+1}^{(l)} }{2} 
        \\
        c_e^{(l+1)} &= \frac{ c_{2e}^{(l)} + c_{2e+1}^{(l)} }{2} 
    \end{align}
\end{subequations}
where $k_e^{(l)}$ and $c_e^{(l)}$ are the material parameters of the $e$\tss{th} element on the $l$\tss{th} level.

This work exclusively considers multigrid V-cycles, the pseudocode of which is presented in Algorithm \ref{alg: stmg v cycle}. In this context, level $l=1$ is the finest level, $N_l$ denotes the number of levels, $\mbf{z}^{(l+1)}$ is the coarse grid correction vector, and $\nu$ denotes the number of pre- and post-smoothing steps. In this paper, $\nu=5$ unless stated otherwise. 

\begin{algorithm}
\caption{$\mathrm{VCycle}(l, \mbf{u}^{(l)}, \mbf{b}^{(l)})$}
\label{alg: stmg v cycle}
\begin{algorithmic}[1]
    \IF{$l = N_l$}
        \STATE Solve for $\mbf{u}^{(l)}$ in the equation $\mbf{J}^{(l)} \mbf{u}^{(l)} = \mbf{b}^{(l)}$
        \RETURN $\mbf{u}^{(l)}$
    \ELSE
        \STATE Perform $\nu$ smoothing steps on $\mbf{J}^{(l)} \mbf{u}^{(l)} = \mbf{b}^{(l)}$
        \STATE $\mbf{b}^{(l+1)} \gets \mbf{R}^{(l)} \left( \mbf{b}^{(l)} - \mbf{J}^{(l)} \mbf{u}^{(l)} \right)$
        \STATE $\mbf{z}^{(l+1)} \gets \mathrm{VCycle}(l+1, \mbf{0}, \mbf{b}^{(l+1)})$
        \STATE $\mbf{u}^{(l)} \gets \mbf{u}^{(l)} + \mbf{P}^{(l)} \mbf{z}^{(l+1)}$
        \STATE Perform $\nu$ smoothing steps on $\mbf{J}^{(l)} \mbf{u}^{(l)} = \mbf{b}^{(l)}$
        \RETURN $\mbf{u}^{(l)}$
    \ENDIF
\end{algorithmic}
\end{algorithm}

\begin{figure}
    \centering
    \includegraphics[width=0.6\linewidth]{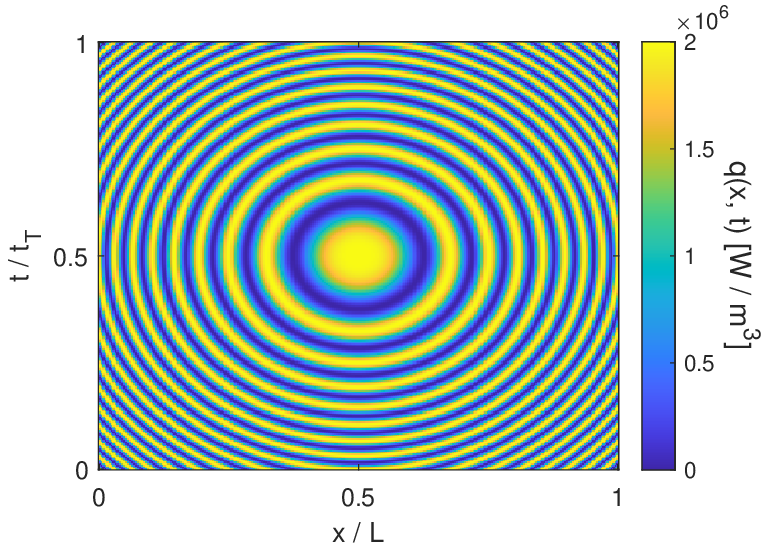}
    \caption{Imposed heat load, $q(x,t)$, used for the test problems considered in this paper, as defined in Equation (\ref{eq: heat load def}). }
    \label{fig: heat load plot}
\end{figure}

The initial guess for the vector of unknowns is $\mbf{u} = \mbf{0}$. The relative residual after $n$ cycles is defined as:
\begin{align}
    r_n &= \frac{\lVert \mbf{J} \mbf{u}_n - \mbf{b} \rVert}{\lVert \mbf{b} \rVert} 
\end{align}
where $\mbf{u}_n$ is $\mbf{u}$ after the $n$\tss{th} cycle and $\lVert \cdot \rVert$ is the Euclidian norm. The multigrid method is terminated after the $n$\tss{th} cycle if the following condition is true:
\begin{align}
    \label{eq: stmg termination criterion}
    r_n < 10^{-9} \text{ or } r_n > 10^{9} \text{ or } n = 100
\end{align}
An estimate of the convergence factor is then computed as:
\begin{align}
    \text{Estimated convergence factor} &= \left( 
    r_N  /  r_0 
    \right)^{1/N}
\end{align}
where $N$ is the number of completed cycles. During testing, it was found that this estimate often depends noticeably on the imposed heat load function, $q(x,t)$. For the sake of consistency, the imposed heat load is set to the following in every test presented in this paper:
\begin{align}
    q_{e,j} &= \left( 1 + \cos \left\{ 200 \cdot \left[ \left( x_e/L - 1/2 \right)^2 + \left( t_j/t_T - 1/2 \right)^2 \right] \right\} \right) \cdot 10^6 \, \mathrm{W/m^3} 
    \label{eq: heat load def}
\end{align}
where $q_{e,j}$ is the heat imposed on the $e$\tss{th} element at $t=t_j$. This function is plotted is Figure \ref{fig: heat load plot}. This is not a realistic heat load in the context of practical heat conduction problems. Instead, the motivation behind the above expression is to expose the multigrid method to an input which contains many different frequencies. This seemed to produce more pessimistic estimates of the convergence factor than, for example, setting $q(x,t)$ to be constant over both $x$ and $t$. 

\subsection{Importance of coarsening type and anisotropy} 

\begin{figure}
    \centering
    \includegraphics[width=0.8\linewidth]{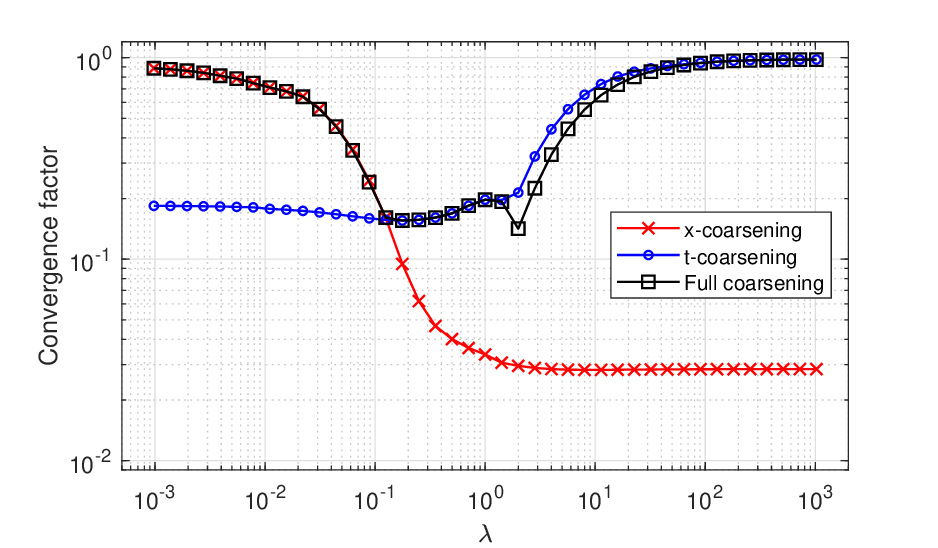}
    \caption{Estimated convergence factors of the space-time multigrid method defined in Section \ref{sec: stmg components def} when using two grids and the material parameters are uniform over space. These are plotted for each coarsening type as a function of $\lambda$, which is defined in Equation (\ref{eq: lambda def}). Specifically, it is the value of $\lambda$ on the finest grid. }
    \label{fig: prob 0 results}
\end{figure}

The performance of STMG methods for diffusion problems are, in general, strongly dependant on the coarsening type and the degree of anisotropy of the discretisation \cite{horton1995_stMultigrid, franco2018_stmg_with_CN, Gander2016_stmg_br}, which is quantified by the following dimensionless ``anisotropy parameter":
\begin{align}
    \label{eq: lambda def}
    \lambda = D \frac{\Delta t}{\Delta x^2}
\end{align}
where $D = k/c$ is the thermal diffusivity. The STMG method studied in this paper is no exception to this rule, as shown in Figure \ref{fig: prob 0 results}. This shows the estimated convergence factors of the two-grid method when applied to a problem where $k=c=1$ at all points, $N_\mathrm{el} = N_t = 256$, $L=1$, and $t_T$ is varied between $2^{-18}$ and $2^2$. This makes it so that $\lambda$ is varied between $2^{-10}$ and $2^{10}$. It is seen in Figure \ref{fig: prob 0 results} that it is optimal to use $x$-coarsening when $\lambda$ is large, and to use $t$-coarsening when $\lambda$ is small. Meanwhile, full space-time coarsening is always suboptimal, and the associated convergence factors are often very close to 1, thus making it impractical. This is consistent with the findings of \cite{horton1995_stMultigrid}. 

The strong dependence on $\lambda$ can be explained by the fact that the stencil in Equation (\ref{eq: stencil of J 1}) can be rewritten as the following:
\begin{align}
    c \frac{\Delta x}{\Delta t} \cdot 
    \left( \,
    \begin{bmatrix}
        0 & 0 & 0 \\ 
        1/6 & 2/3 & 1/6 \\
        -1/6 & -2/3 & -1/6
    \end{bmatrix}
    +
    \lambda \cdot
    \begin{bmatrix}
        0 & 0 & 0 \\ 
        -1 & 2 & -1 \\
        0 & 0 & 0
    \end{bmatrix}
    \, \right)
\end{align}
Here it can be seen that the spatial coupling becomes dominant when $\lambda \gg 1$, while the temporal coupling becomes dominant when $\lambda \ll 1$. Pointwise smoothers mainly reduce components of the error which oscillate in the direction of strongest coupling \cite{horton1995_stMultigrid}, which explains the behaviour seen in Figure \ref{fig: prob 0 results}. There are other types of smoothers which reduce error oscillations in both the $x$- and $t$-directions, including block-Jacobi smoothers \cite{Gander2016_stmg_br} and zebra smoothers \cite{franco2018_stmg_with_CN}, which make it viable to use full space-time coarsening. However, these smoothers are more difficult to implement in parallel compared to pointwise smoothers. Additionally, STMG methods using block-Jacobi smoothers have been reported to be slower than ones using pointwise smoothers due to greater computational overhead \cite{falgout2017_stmgComparisons}. As such, this paper will not consider full space-time coarsening any further. 


The above observations form the motivation for the use of adaptive coarsening strategies, like the one in Algorithm \ref{alg: coar strat for uniform}, which was originally proposed in \cite{horton1995_stMultigrid}. In this strategy, the value of $\lambda$ is computed on each level, and these values are then used to decide if the next coarse level should be constructed using $x$- or $t$-coarsening. These decisions are made by comparing $\lambda$ to a constant denoted $\lambda_\mathrm{crit}$, which is the threshold where the convergence factors for $x$-coarsening and $t$-coarsening are equal. For example, in Figure \ref{fig: prob 0 results} it is seen that $\lambda_\mathrm{crit} \approx 0.1$ for the considered method, since this is the point where the curves intersect. The value of $\lambda_\mathrm{crit}$ varies between different discretisations and STMG methods, but it often lies between 0.1 and 1.0 \cite{horton1995_stMultigrid, franco2018_stmg_with_CN, Gander2016_stmg_br}.

\begin{algorithm}
\caption{Coarsening strategy for uniform material parameters.}
\label{alg: coar strat for uniform}
\begin{algorithmic}[1]
    \STATE Construct level 1 (the finest level)
    \FOR {$l = 1, \hdots, N_l-1$}
        \STATE $\lambda^{(l)} \gets D \dfrac{\Delta t^{(l)}}{  \left(\Delta x^{(l)}\right)^2  }$
        \IF{$\lambda^{(l)} < \lambda_\mathrm{crit}$}
            \STATE Construct level $l+1$ by coarsening level $l$ in the $t$-direction
        \ELSE
            \STATE Construct level $l+1$ by coarsening level $l$ in the $x$-direction
        \ENDIF
    \ENDFOR
\end{algorithmic}
\end{algorithm}

\subsection{Problem statement} 
\label{sec: problem statement}

In cases where $k$ and $c$ are variable over space, the diffusivity of each element, $e$, can be defined as:
\begin{align}
    D_e \equiv k_e/c_e
\end{align}
This, in turn, means that the anisotropy parameter of each element can be defined as: 
\begin{align}
    \lambda_e \equiv D_e \frac{\Delta t}{\Delta x^2}
\end{align}
For realistic problems involving both thermal conductors and insulators, the contrast in diffusivity can easily be on the order of $10^3$. For uniform $\Delta t$ and $\Delta x$, this implies that the contrast in $\lambda_e$ can also be on the order of $10^3$, meaning that $\lambda_e$ can simultaneously be significantly greater than and less than $\lambda_\mathrm{crit}$ in different parts of the domain. 
This raises the question: Which coarsening strategy should be used in such cases with high contrast in the diffusivity? Also, in the context of topology optimisation of heat-conducting structures, the structures which are generated often contain very small features in the form of thin branches \cite{Alexandersen2016_3dnatconv, zhou2025robustsolverlargescaleheat} which will not be properly resolved on the coarse grids if too much $x$-coarsening is applied. This raises the question: Is it beneficial to replace some of the $x$-coarsening steps in the coarsening strategy with $t$-coarsening steps when dealing with small features? This paper tries to answer these two questions under the constraint of using only Geometric MultiGrid (GMG) methods and uniform Cartesian space-time meshes. 


\section{Effective anisotropy parameter} 
\label{sec: eff lambda}

Through trial and error it was found that the following ``effective anisotropy parameter" works well as an indicator for when to use $x$- or $t$-coarsening when considering high-contrast problems:
\begin{align}
    \label{eq: lambda eff def 1}
    \lambda_\mathrm{eff} &= \sqrt{ \min_e(\lambda_e) \max_e(\lambda_e) }
\end{align}
An equivalent way to define $\lambda_\mathrm{eff}$ is as follows:
\begin{align}
    \lambda_\mathrm{eff} &= D_\mathrm{eff} \frac{\Delta t}{\Delta x^2}
\end{align}
where $D_\mathrm{eff}$ is the ``effective diffusivity", which is the following:
\begin{align}
    D_\mathrm{eff} &= \sqrt{ \min_e(D_e) \max_e(D_e) } 
\end{align}
This is the geometric mean of the minimum and maximum diffusivity. 

\begin{figure}
    \centering
    \includegraphics[width=0.7\linewidth]{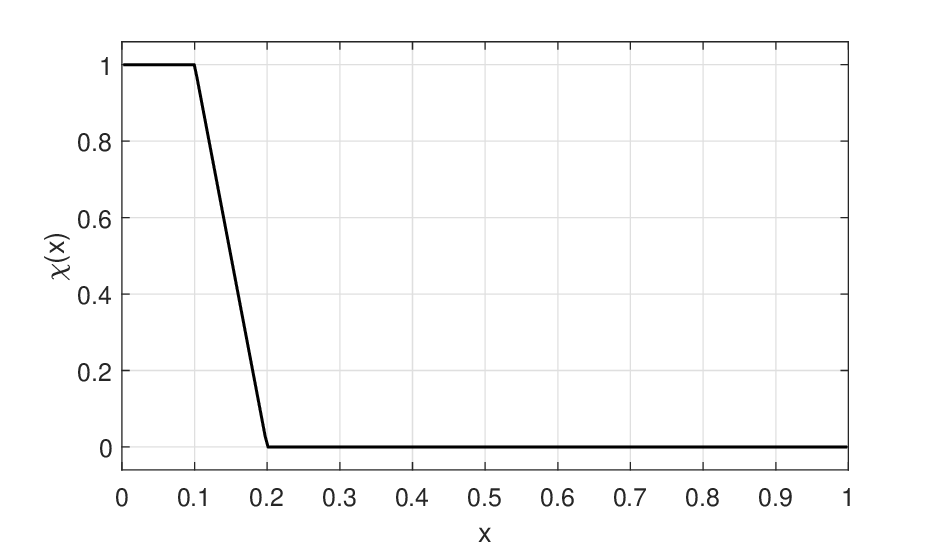}
    \caption{Design field, $\chi$, defined for problem 3 in Section \ref{sec: eff lambda}. }
    \label{fig: prob 3 des field}
\end{figure}

\begin{figure}
    \centering
    \subfloat[Problem 1.]{\includegraphics[width=0.4\textwidth]{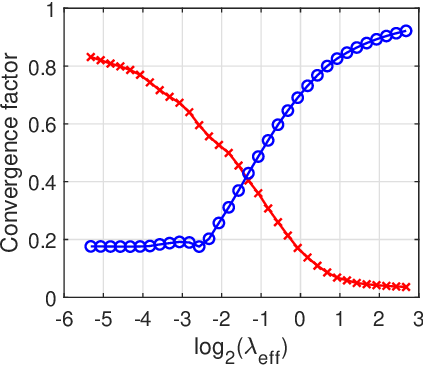}} \hspace{5mm}
    \subfloat[Problem 2.]{\includegraphics[width=0.4\textwidth]{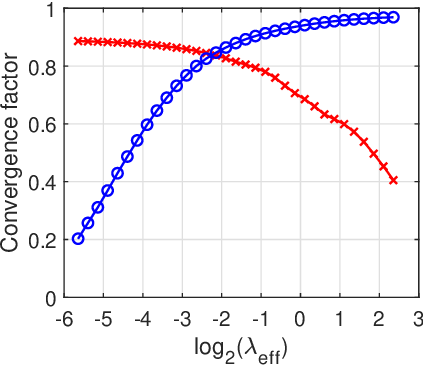}}
    
    \subfloat[Problem 3.]{\includegraphics[width=0.4\textwidth]{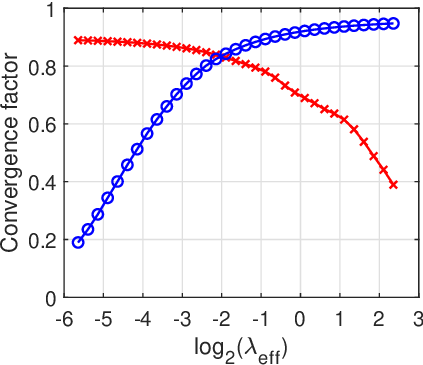}} \hspace{5mm}
    \subfloat[Problem 4.]{\includegraphics[width=0.4\textwidth]{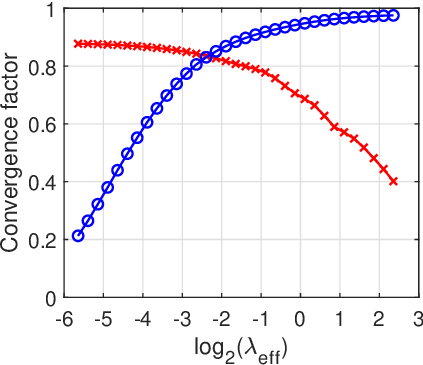}}
    
    \subfloat[Problem 5.]{\includegraphics[width=0.4\textwidth]{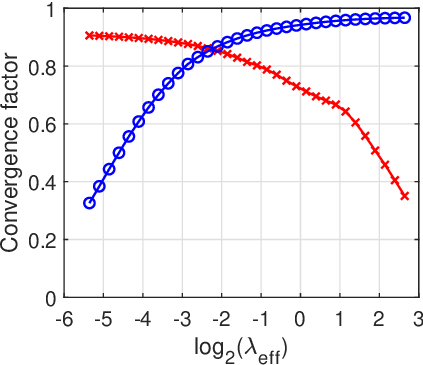}} \hspace{5mm}
    \subfloat[Problem 6.]{\includegraphics[width=0.4\textwidth]{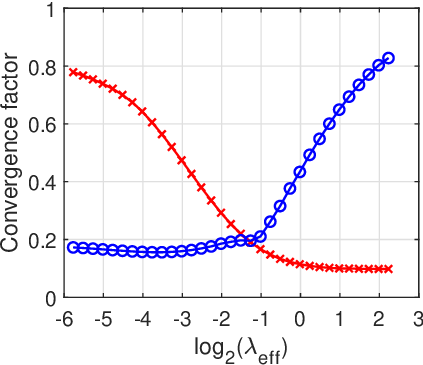}}
    \caption{Estimated convergence factors for the two-grid space-time multigrid method applied to the six test problems considered in Section \ref{sec: eff lambda}. The red crossed curves show the convergence factors when using $x$-coarsening, and the blue circled curves show the convergence factors when using $t$-coarsening. These are plotted as functions of $\log_2 (\lambda_\mathrm{eff} )$ on the finest level, where $\lambda_\mathrm{eff}$ is the quantity defined in Equation (\ref{eq: lambda eff def 1}).}
    \label{fig: prob 1 to 6 results}
\end{figure}

To demonstrate that $\lambda_\mathrm{eff}$ works well as an indicator for which coarsening type to choose, a series of test problems with moderate/high contrast in the material parameters are considered. In each problem, the design field, $\chi$, on the $e$\tss{th} element is of the following form:
\begin{align}
    \label{eq: prob 1-6 design field}
    \chi_e &= f_{01}\left[ 1/2 - \alpha (x_e - x_\mathrm{offset}) \right]
\end{align}
where $\alpha$ and $x_\mathrm{offset}$ are constants and $f_{01}$ is a function which makes sure that $\chi$ is always between 0 and 1. This function is defined as:
\begin{align}
    f_{01}(\chi) &= \begin{cases}
        0 & \text{if } \chi \leq 0 \\ 
        \chi & \text{if } 0 < \chi < 1 \\
        1 & \text{if } \chi \geq 1
    \end{cases}
\end{align}
These design fields are piecewise linear functions consisting of three pieces: 
\begin{itemize}
    \item A region at $x \leq x_\mathrm{offset} - 1/(2\alpha)$ which is filled with the conductor. 
    \item A region at $x \geq x_\mathrm{offset} + 1/(2\alpha)$ which is filled with the insulator. 
    \item A transitional region centred at $x = x_\mathrm{offset}$ with a width of $1/\alpha$ where $\chi$ drops from 1 to 0.
\end{itemize}
Figure \ref{fig: prob 3 des field} shows an example of such a design field. 
The problems are non-dimensionalised by defining the following scales for the variables: the length scale is set to $L$, the conductivity scale is set to $k_\mathrm{con}$, the heat capacity scale is set to $c_\mathrm{con}$, and the time scale is set to $L^2 c_\mathrm{con} / k_\mathrm{con}$. This makes it so that the system only has the following free parameters:
\begin{itemize}
    \item $k_\mathrm{ins}$ and $c_\mathrm{ins}$, which quantify the contrasts in $k$ and $c$;
    \item $\alpha$, which is set to 10 to avoid situations where the performance is worsened due to poor resolution of the transitional region;
    \item $x_\mathrm{offset}$, which is approximately equal to the fraction of the domain which is filled with the conductor;
    \item $N_\mathrm{el}$ and $N_t$, which are both set to 256;
    \item $t_T$, which is varied for the purpose of varying $\lambda_\mathrm{eff}$. Specifically, $t_T$ is set to powers of $2^{1/4}$ which are chosen such that $2^{-6} < \lambda_\mathrm{eff} < 2^3$. 
\end{itemize}
Six different problems are considered, the parameters of which are listed in Table \ref{tab: def of problem 1-6}, along with the considered ranges of $t_T$. The purposes of these problems are the following:
\begin{itemize}
    \item Problem 1 and 2 demonstrate that $\lambda_\mathrm{eff}$ works reliably for both moderate ($10^2$) and high ($10^4$) contrast in $k$. 
    \item Problem 2, 3, and 4 demonstrate that $\lambda_\mathrm{eff}$ works reliably regardless of whether there are equal or unequal amounts of each material. 
    \item Problem 5 demonstrates that $\lambda_\mathrm{eff}$ works reliably when there is only high contrast in $c$. 
    \item Problem 6 demonstrates that $\lambda_\mathrm{eff}$ works reliably when there is high contrast in both $c$ and $k$. This problem is also noteworthy because the conductor and insulator have equal thermal diffusivities, while the diffusivity in the transitional region is lower due to the interpolations defined in Equations (\ref{eq: SIMP k}) and (\ref{eq: SIMP c}). 
\end{itemize}
\begin{table}
\centering
\begin{tabular}{ccccc}
\hline
Problem number & $k_\mathrm{ins}$ & $c_\mathrm{ins}$ & $x_\mathrm{offset}$ & Range of $t_T$ \\ \hline

1 & $10^{-2}$ & $1$ & $0.5$ & $2^{-10}$ to $2^{-2}$ \\  
2 & $10^{-4}$ & $1$ & $0.5$ & $2^{-7}$ to $2^{1}$ \\ 
3 & $10^{-4}$ & $1$ & $0.15$ & $2^{-7}$ to $2^{1}$ \\ 
4 & $10^{-4}$ & $1$ & $0.85$ & $2^{-7}$ to $2^{1}$ \\ 
5 & $1$ & $10^{-4}$ & $0.5$ & $2^{-20}$ to $2^{-12}$ \\ 
6 & $10^{-4}$ & $10^{-4}$ & $0.5$ & $2^{-12}$ to $2^{-4}$ \\ \hline

\end{tabular}
\caption{Parameter values of the six test problems considered in Section \ref{sec: eff lambda}. }
\label{tab: def of problem 1-6}
\end{table}
The considered two-grid STMG method is applied to all six problems, and Figure \ref{fig: prob 1 to 6 results} shows the results. It is seen that the intersections between the $x$-coarsening and $t$-coarsening curves are in the range $2^{-3} \leq \lambda_\mathrm{eff} \leq 2^{-1}$ for all six problems. This is a relatively narrow range of values of $\lambda_\mathrm{eff}$, considering that the contrast in the diffusivity was $10^4$ (which is $\approx 2^{13.3}$) in most of the test problems. As such, $\lambda_\mathrm{eff}$ is deemed to be a reasonably reliable indicator. Other expressions for $\lambda_\mathrm{eff}$ were also tested as part of this work. None of those expressions were as reliable as the one in Equation (\ref{eq: lambda eff def 1}), as the intersection points were found in a much broader range of values of $\lambda_\mathrm{eff}$. All the tested expressions are listed in Table \ref{tab: tested exprs for lambda eff}, along with a short summary of the findings of the tests.

\bgroup
\def\arraystretch{1.08}
\begin{table}
\centering
\begin{tabular}{ccc}
\hline
\textbf{Expression for $\lambda_\mathrm{eff}$}  & \textbf{\begin{tabular}[c]{@{}l@{}}Works well for equal\\amounts of each material?\end{tabular}} & \textbf{\begin{tabular}[c]{@{}l@{}}Works well for unequal\\amounts of each material?\end{tabular}}  \\ \hline 
$ \sqrt{ \min_e(\lambda_e) \max_e(\lambda_e) } $ & Yes & Yes \\ 
$ \left( \prod_e \lambda_e \right)^{1/N_\mathrm{el}} $ & Yes & No \\   
$ \left( \min_e(\lambda_e) + \max_e(\lambda_e) \right) / \, 2 $ & No & No \\ 
$ \left( \sum_e \lambda_e \right) / \, N_\mathrm{el} $ & No & No \\ 
$ 2 \, / \left( \min_e(\lambda_e)^{-1} + \max_e(\lambda_e)^{-1} \right) $ &  No & No  \\ 
$ N_\mathrm{el} \, / \left( \sum_e \left( {\lambda_e}^{-1} \right) \right) $ &  No & No  \\     
$ \min_e(\lambda_e) $ &  No & No  \\ 
$ \max_e(\lambda_e) $ &  No & No  \\ \hline

\end{tabular}
\caption{List of expressions that were considered as candidates for how to define the effective anisotropy parameter, $\lambda_\mathrm{eff}$, along with the findings of the tests of each of these expressions. There is only one expression which works well for both equal and unequal amounts of each material, which is the one at the top of the list. As such, this expression was adopted as being the definition of $\lambda_\mathrm{eff}$. }
\label{tab: tested exprs for lambda eff}
\end{table}
\egroup

Based on these findings, a coarsening strategy based on $\lambda_\mathrm{eff}$ is defined in Algorithm \ref{alg: coar strat for high contrast}. This strategy is identical to Algorithm \ref{alg: coar strat for uniform}, except it computes $D_\mathrm{eff}$ and $\lambda_\mathrm{eff}$ on each level, and $\lambda$ is replaced with $\lambda_\mathrm{eff}$. This strategy will be used in the next section. 

\begin{algorithm}
\caption{Coarsening strategy for high-contrast problems.}
\label{alg: coar strat for high contrast}
\begin{algorithmic}[1]
    \STATE Construct level 1 (the finest level)
    \FOR {$l = 1, \hdots, N_l-1$}
        \STATE $D_\mathrm{eff}^{(l)} \gets \sqrt{ \min_e\left(D_e^{(l)}\right) \max_e\left(D_e^{(l)}\right) } $
        \STATE $\lambda_\mathrm{eff}^{(l)} \gets D_\mathrm{eff}^{(l)} \dfrac{\Delta t^{(l)}}{  \left(\Delta x^{(l)}\right)^2  }$
        \IF{$\lambda_\mathrm{eff}^{(l)} < \lambda_\mathrm{crit}$}
            \STATE Construct level $l+1$ by coarsening level $l$ in the $t$-direction
        \ELSE
            \STATE Construct level $l+1$ by coarsening level $l$ in the $x$-direction
        \ENDIF
    \ENDFOR
\end{algorithmic}
\end{algorithm}

The proposed coarsening strategy differs from the coarsening strategies used by \citet{schwalsberger2018_eddyThesis} and \citet{neumuller2019_stmgEddy}. In both of these works, the coarsening type is chosen based on quantities which are analogous to $\min_e(\lambda_e)$, as opposed to the effective anisotropy parameter defined in Equation (\ref{eq: lambda eff def 1}). This makes sense in the context of their work, because they used specialised smoothers which make it so that the convergence factor of the STMG decreases when $\lambda$ is increased. Therefore, it makes sense that the convergence factors would be limited by the minimum of $\lambda$. Based on this, it can be said that the coarsening strategy in Algorithm \ref{alg: coar strat for high contrast} will not necessarily be optimal for all choices of smoothers or other STMG components. Therefore, if the reader wishes to implement other STMG methods for high-contrast problems, then the reader is encouraged to check if other definitions of $\lambda_\mathrm{eff}$ work better in the considered context.

\section{Different rediscretisation methods} 
\label{sec: different redisc methods}

So far, this paper has presented one method of rediscretising the coarse levels of the space-time multigrid method. This section will present the other rediscretisation methods which were tested as part of this work.

\subsection{Definitions of rediscretisation methods}
\label{sec: different redisc methods defs}

In this context, the term ``rediscretisation method" refers to a pair of the following two components: an interpolation method which is used to define the prolongation (and restriction) operators, and a method of reassembling the system matrices on the coarse levels. Two interpolation methods are considered, each of which are assigned a one-letter abbreviation. These methods are the following:
\begin{itemize}
    \item {Causal interpolation:} (abbreviated C)
    
    This is the interpolation method which was defined in Section \ref{sec: stmg components def}. For this method, the stencils for the prolongation operators for $x$- and $t$-coarsening are:
    \begin{align}
        x\text{-coarsening: } 
        \frac{1}{2} 
        \left] \begin{matrix}
            0 & 0 & 0 \\
            1 & 2 & 1 \\
            0 & 0 & 0
        \end{matrix} \right[ 
        \quad
        t\text{-coarsening: } 
        \left] \begin{matrix}
            0 & 1 & 0 \\
            0 & 1 & 0 \\
            0 & 0 & 0
        \end{matrix} \right[ 
    \end{align}
    
    \item {Bilinear space-time interpolation:} (abbreviated B)
    
    For this method, the stencils for the prolongation operators for $x$- and $t$-coarsening are:
    \begin{align}
        x\text{-coarsening: } 
        \frac{1}{2} 
        \left] \begin{matrix}
            0 & 0 & 0 \\
            1 & 2 & 1 \\
            0 & 0 & 0
        \end{matrix} \right[ 
        \quad
        t\text{-coarsening: } 
        \frac{1}{2} 
        \left] \begin{matrix}
            0 & 1 & 0 \\
            0 & 2 & 0 \\
            0 & 1 & 0
        \end{matrix} \right[ 
    \end{align}
\end{itemize}
Four reassembly methods are considered, each of which are assigned a one-letter abbreviation. These methods are the following:
\begin{itemize}

    \item {Averaging of conductivity:} (abbreviated K)

    This is the reassembly method which was defined in Section \ref{sec: stmg components def}. This method reassembles the matrices according to the BE-FE discretisation method, and when it applies $x$-coarsening, $k$ and $c$ on the coarse levels are computed by averaging $k$ and $c$ on the finer levels like so:
    \begin{subequations}
    \label{eq: K method def}
    \begin{align}
        k_e^{(l+1)} &= \frac{ k_{2e}^{(l)} + k_{2e+1}^{(l)} }{2} 
        \\
        \label{eq: coarse c in K method def}
        c_e^{(l+1)} &= \frac{ c_{2e}^{(l)} + c_{2e+1}^{(l)} }{2} 
    \end{align}
    \end{subequations}
    
    \item {Averaging of design field:} (abbreviated D)

    This method also reassembles the matrices according to the BE-FE discretisation method, and when applying $x$-coarsening, the design field, $\chi$, on the coarse levels is defined by averaging $\chi$ on the finer levels:
    \begin{align}
        \chi_e^{(l+1)} &= \frac{ \chi_{2e}^{(l)} + \chi_{2e+1}^{(l)} }{2} 
    \end{align}
    The conductivity and heat capacity on the coarse levels are then defined using the SIMP scheme:
    \begin{subequations}
    \begin{align}
        k_e^{(l+1)} &= k_\mathrm{SIMP} \left( \chi_e^{(l+1)} \right)
        \\
        c_e^{(l+1)} &= c_\mathrm{SIMP} \left( \chi_e^{(l+1)} \right)
    \end{align}
    \end{subequations}
    Note that this method is only defined in the context of density-based topology optimisation, since it relies on the presence of the design field and penalisation scheme. 
    
    \item {Averaging of resistivity:} (abbreviated R)
    
    This method reassembles the matrices according to the BE-FE discretisation method. For this method, the thermal resistivity, $\rho$, on each element is defined as:
    \begin{align}
        \rho_e^{(l)} \equiv 1/ k_e^{(l)}
    \end{align}
    When applying $x$-coarsening, $\rho$ on the coarse levels is computed by averaging $\rho$ on the finer levels:
    \begin{align}
        \rho_e^{(l+1)} &= \frac{ \rho_{2e}^{(l)} + \rho_{2e+1}^{(l)} }{2} 
    \end{align}
    The conductivity is then computed based on $\rho$
    \begin{align}
        k_e^{(l+1)} &= 1 / \rho_e^{(l+1)}
    \end{align}
    after which the heat capacity is computed according to Equation (\ref{eq: coarse c in K method def}). An equivalent way to express the conductivity on level $l+1$ is as follows:
    \begin{align}
        k_e^{(l+1)} &= \frac{2 k_{2e}^{(l)} k_{2e+1}^{(l)} }{ k_{2e}^{(l)} + k_{2e+1}^{(l)} }
    \end{align}
    This way of defining $k_e^{(l+1)}$ is a one-dimensional analogue of the rediscretisation method proposed for electrical conductivity in \cite{Moucha2004_anisoCondMG}. 
    
    \item {Projection of system matrix:} (abbreviated P) 
    
    In this method, the system matrices on the coarse levels are computed by projecting the system matrices on the finer levels like so:
    \begin{align}
        \mbf{J}^{(l+1)} &= \mbf{R}^{(l)} \, \mbf{J}^{(l)} \, \mbf{P}^{(l)}
    \end{align}
    This is also referred to as Galerkin projection \cite{frederickson1988_tpma, amir2014_topOptMG} or the Galerkin approximation \cite{wesseling1982}. This reassembly method presents a challenge in the context of the proposed coarsening strategy, because this method does not explicitly return the values of $k_e$ and $c_e$ on the coarse levels, which are needed to compute $\lambda_\mathrm{eff}$. In theory, it might be possible to extract the anisotropy parameter from the coefficients in the system matrices. However, it is noted that the projection method is similar to the reassembly method where $k$ and $c$ are averaged (the K-method), since both of these methods produce linear combinations of the coefficients in the PDE. For this reason, the values of $k_e$ and $c_e$ on the coarse levels are approximated using Equation (\ref{eq: K method def}) when applying $x$-coarsening, and these approximations are then used to compute $\lambda_\mathrm{eff}$. 
\end{itemize}
Each of the two interpolation methods can be combined with any of the four reassembly methods, so there are a total of eight different rediscretisation methods which can be constructed from the above methods. These rediscretisation methods are named by combining the abbreviations of the corresponding interpolation and reassembly methods. For example, the rediscretisation method defined in Section \ref{sec: stmg components def} is given the identifier ``CK".

\subsection{Tests of different rediscretisation methods}
\label{sec: different redisc methods tests}

\begin{figure}
    \centering
    \includegraphics[width=0.8\linewidth]{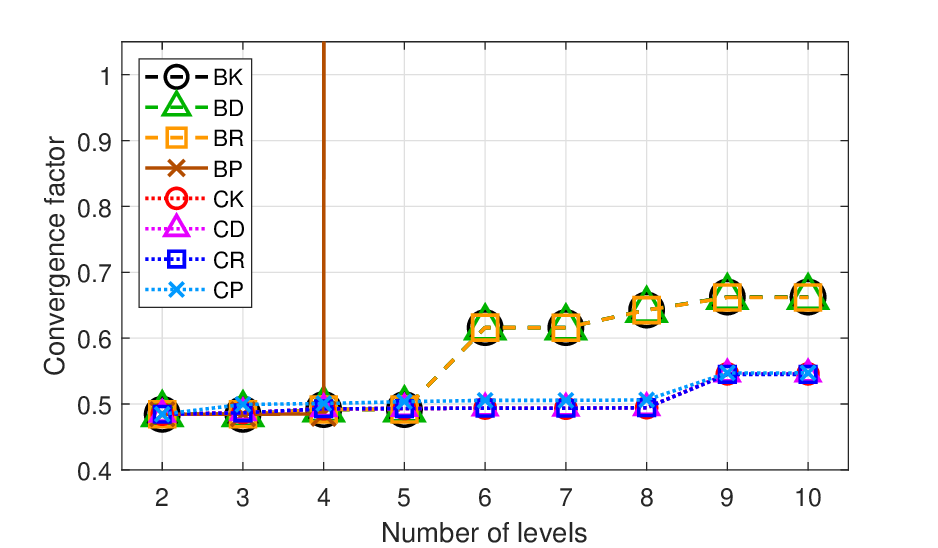}
    \caption{Results of the tests of different rediscretisation methods applied to problem 7 as defined in Section \ref{sec: different redisc methods tests}, which is a problem where all spatial features are easy to resolve. These rediscretisation methods are defined in Section \ref{sec: different redisc methods defs}. The vertical line at 4 levels is the curve for the BP-method, which becomes unstable when the number of levels is greater than 4 for this problem. }
    \label{fig: prob 7 results}
\end{figure}

To test the proposed rediscretisation methods, three more test problems are defined. For these problems, $L=0.1\,\mathrm{m}$, $N_\mathrm{el}=N_t=256$, and the material parameters of the conductor and insulator are listed in Table \ref{tab: prob 7-9 thermal properties}. 
\begin{table}[htbp]
\centering
\begin{tabular}{lrr}
\hline
Material & $k \, [\mathrm{W \, m^{-1} \, K^{-1}}]$ & $c \, [\mathrm{J \, m^{-3} \, K^{-1}}]$ \\ \hline
Conductor (aluminium)           & $2.14 \times 10^2$    & $2.41 \times 10^6$  \\ 
Insulator (epoxy resin polymer) & $1.97 \times 10^{-1}$ & $1.67 \times 10^6$ \\ 
\hline
\end{tabular}
\caption{Material properties for the test problems considered in Section \ref{sec: different redisc methods} and \ref{sec: new coar strat}. These are based on measurements in \cite{topOptExperiment}. }
\label{tab: prob 7-9 thermal properties}
\end{table}
These are the measured thermal properties of aluminium and a specific epoxy resin polymer as stated in \cite{topOptExperiment}, except rounded to three significant digits. The individual test problems are:
\begin{itemize}

    \item Problem 7, where the design field is defined such that all the spatial features are easy to resolve, even on very coarse meshes. Specifically, it is given by Equation (\ref{eq: prob 1-6 design field}) with $\alpha=5/L$ and $x_\mathrm{offset}=L/2$. The terminal time is $t_T = 10 \,\mathrm{s}$, which means that $\lambda_\mathrm{eff} \approx 0.83$ on the finest level. This makes it so that the coarsening strategy in Algorithm \ref{alg: coar strat for high contrast} will apply a mix of $x$-coarsening and $t$-coarsening on the finer levels. 

    \item Problem 8, where the domain consists of two conducting regions separated by a thin insulating region. Specifically, the design field is given by the following:
    \begin{align}
        \label{eq: problem 8 chi def}
        \chi_e &= f_{01}\left( \left| \frac{x_e - L/2}{L} \right| \cdot 2/F - 1 \right) 
    \end{align}
    where $F=0.03$ is the fraction of the domain which is filled purely with the insulator. In addition, the fraction of the domain where $0<\chi<1$ is also equal to $F$. This design field is plotted in Figure \ref{fig: prob 8 des field}, except $F$ is set to $0.1$ for illustrative purposes. The terminal time is $t_T = 100 \,\mathrm{s}$, which means that $\lambda_\mathrm{eff} \approx 8.3$ on the finest level. This makes it so that the coarsening strategy will apply $x$-coarsening on many of the finer levels. As such, this problem serves as a test for how well the methods perform when there are small spatial features which are poorly resolved on the coarse grids. 

    \begin{figure}
        \centering
        \includegraphics[width=0.7\linewidth]{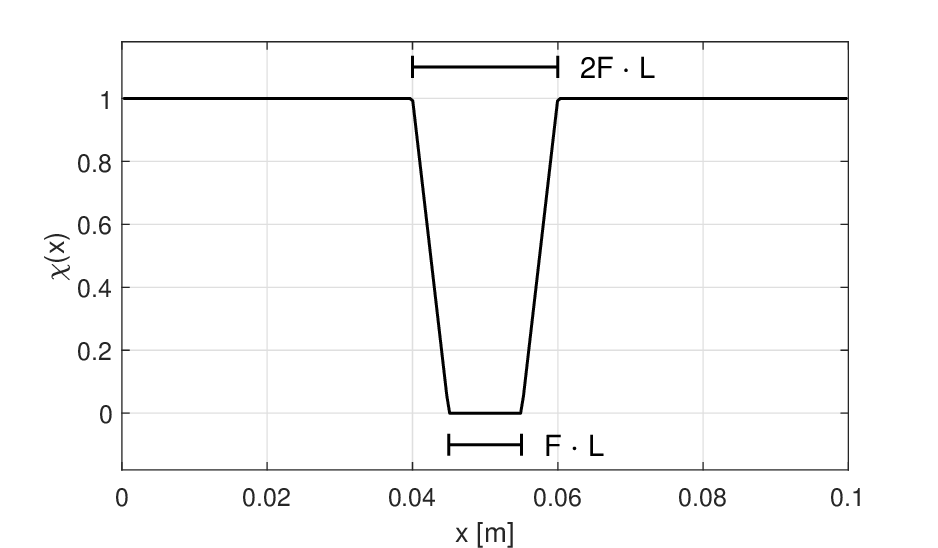}
        \caption{The design field, $\chi$, which is defined in Equation (\ref{eq: problem 8 chi def}), with $L=0.1\,\mathrm{m}$ and $F=0.1$. }
        \label{fig: prob 8 des field}
    \end{figure}

    \item Problem 9, where the design field is the following:
    \begin{align}
        \chi_e &= 1 - f_{01}\left( \left| \frac{x_e - L/2}{L} \right| \cdot 2/F - 1 \right) 
    \end{align}
    This is the exact reverse of problem 8, meaning that the domain consists of two insulating regions separated by a thin conducting region. For this problem, $F=0.03$ and $t_T = 100 \,\mathrm{s}$, like in problem 8. 
    
\end{itemize}
The considered STMG method is applied to the above test problems where $N_l$ (the number of levels) is varied between 2 and 10. For this, the coarsening strategy in Algorithm \ref{alg: coar strat for high contrast} is used, and the parameter $\lambda_\mathrm{crit}$ is set to 0.25. The results for problem 7 are shown in Figure \ref{fig: prob 7 results}. Here it is seen that there is one method which behaves very differently from the others, namely the BP-method, for which the convergence factor explodes after $N_l=4$. Specifically, the convergence factor grows to $\approx 8 \times 10^5$ at $N_l=5$, after which it continues growing exponentially to $\approx 4 \times 10^{31}$ at $N_l=10$. This instability is likely due to the smoother becoming unstable on the coarse levels, as it was found that the convergence factors become even larger when the number of smoothing steps is increased. Similar instability is also observed for the BP method in problem 8 and 9. For problem 8, the instability appears at $N_l=10$, and for problem 9 it appears at $N_l \geq 7$. As such, the BP method will not be considered any further.

Aside from the BP method, the remaining seven methods are stable. For $N_l \leq 5$, the performance of these methods are nearly identical, while for $N_l \geq 6$, the methods using bilinear interpolation become noticeably worse than those using causal interpolation. It is also noted that the choice of reassembly method only has a minor impact on the convergence factors.

\begin{figure}
    \centering
    \includegraphics[width=0.8\linewidth]{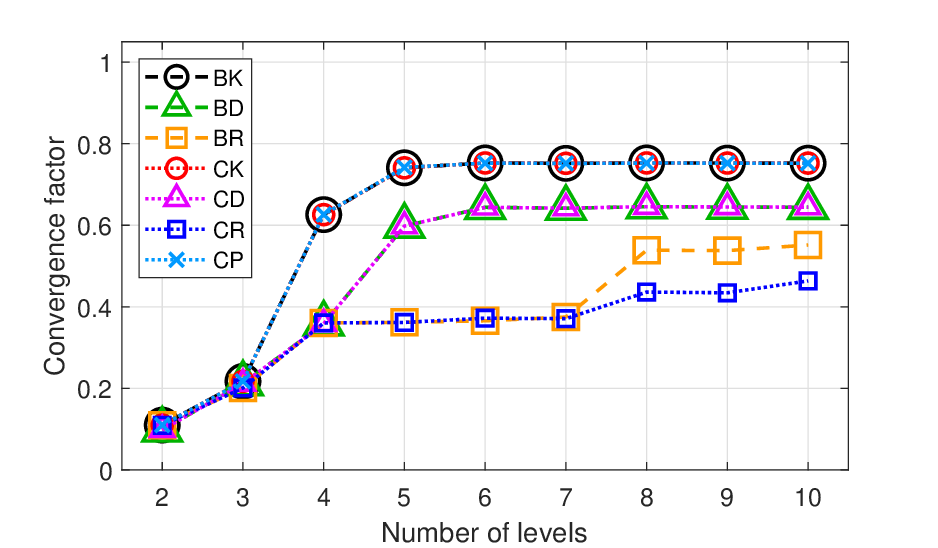}
    \caption{Results of the tests of different rediscretisation methods applied to problem 8 as defined in Section \ref{sec: different redisc methods tests}, which considers two conducting regions separated by a thin insulating region. These rediscretisation methods are defined in Section \ref{sec: different redisc methods defs}. The rediscretisation method named BP is omitted from this plot because it was found to be unstable. }
    \label{fig: prob 8 results}
\end{figure}

The results for problem 8 are shown in Figure \ref{fig: prob 8 results}. Unlike in problem 7, it is seen that the choice of matrix reassembly method has a major impact on the performance, at least for $N_l \geq 4$. The R-method seems to perform the best, and the D-method seems to be second best. The K-method and P-method seem to have similar performance to each other, and they perform the worst. In most cases, the choice of interpolation method only has a minor impact on the performance, unlike in problem 7. However, there are some notable exceptions, namely for the BR- and CR-method at $N_l \geq 8$, where causal interpolation is noticeably better than bilinear interpolation, which is consistent with the results of problem 7. 

\begin{figure}
    \centering
    \includegraphics[width=0.8\linewidth]{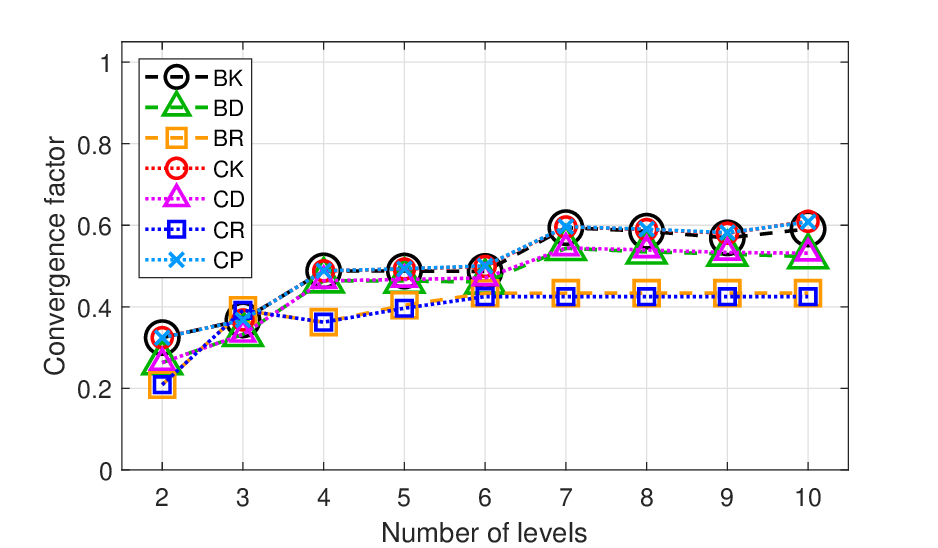}
    \caption{Results of the tests of different rediscretisation methods applied to problem 9 as defined in Section \ref{sec: different redisc methods tests}, which considers two insulating regions separated by a thin conducting region. These rediscretisation methods are defined in Section \ref{sec: different redisc methods defs}. The rediscretisation method named BP is omitted from this plot because it was found to be unstable. }
    \label{fig: prob 9 results}
\end{figure}

The results for problem 9 are shown in Figure \ref{fig: prob 9 results}. This shows the same general trends that were observed for problem 8, namely that the R-method is best, the D-method is second best, and the K- and P-methods are the worst. However, the differences between the methods are less pronounced in this problem. Also, it is noted that there are some cases where bilinear interpolation performs better than causal interpolation, especially around $N_l \approx 10$. However, the impact of the choice of interpolation methods is very minor for this problem. 

\subsection{Summary and discussion regarding rediscretisation methods}
\label{sec: redisc methods discussion}

To summarise the findings regarding the rediscretisation methods, it can be said that bilinear interpolation should not be combined with projection of the system matrices, since this makes the considered STMG method unstable. Aside from that, it can be said that the interpolation method seems to be the dominant deciding factor for the performance when considering problems without small features. Specifically, causal interpolation performs better than bilinear interpolation. However, for problems with small features, the matrix reassembly method is the dominant factor. Specifically, the method which averages the resistivity (the R-method) seems to be the best of the methods which were tested. 

It is argued that it makes sense that the R-method is observed to be the best reassembly method when applied to the above problems. This is because this paper only considers heat conduction in one spatial dimension, which means that the simulation domain can be interpreted as being a set of thermal resistances which are connected to each other in series. As such, the thermal resistances combine additively if they are lumped together, which effectively means that the resistivities are averaged, which is exactly how the R-method processes the resistivities. 

However, if the multigrid method is applied to heat conduction problems with two or more spatial dimensions, then there are many cases where it will make more sense to interpret the heat as flowing through a set of thermal resistances which are connected to each other in parallel. In particular, for commercial heat sinks consisting of fins and tree-like structures generated by topology optimisation methods, it probably makes more sense to think of the heat as flowing in parallel through the fins and branches. If so, then the thermal conductances, not the thermal resistances, should combine additively when lumped together. This means that the conductivities should be averaged, so it would make sense to use the K-method for those problems. The P-method will probably also work well, since it was argued in Section \ref{sec: different redisc methods defs} that the K- and P-methods work similarly to each other.

\section{Coarsening strategy ensuring spatial resolution on coarse grids} 
\label{sec: new coar strat}

\begin{figure}
    \centering
    \includegraphics[width=0.55\linewidth]{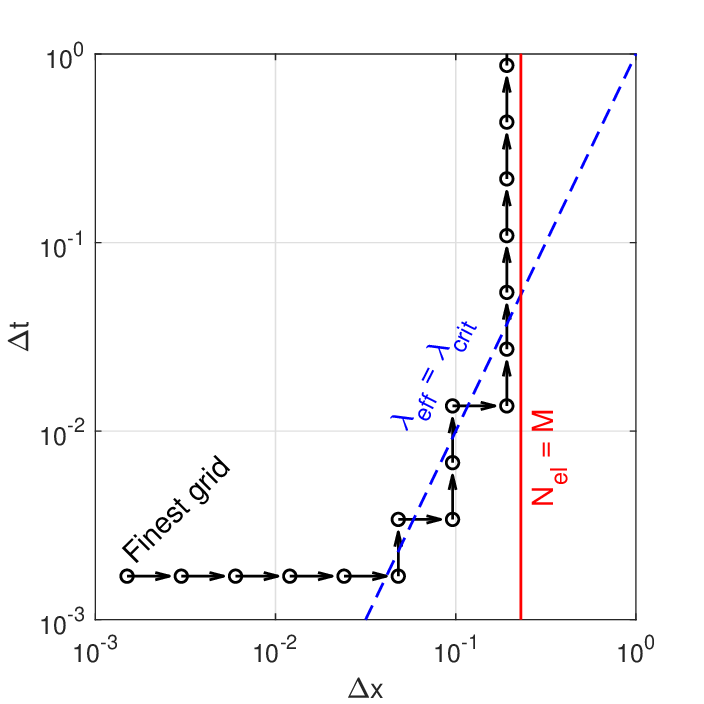}
    \caption{Example of a coarsening path which can be chosen by the coarsening strategy in Algorithm \ref{alg: coar strat for small features}. Each circle represents a grid of the space-time multigrid method, and the arrows show the coarsening direction chosen for each grid. In this example, the strategy first applies $x$-coarsening to many of the finest levels, then zigzags around the dashed line where $\lambda_\mathrm{eff} = \lambda_\mathrm{crit}$, and then avoids crossing the solid line where $N_\mathrm{el} = M$ by applying $t$-coarsening to the coarsest levels. }
    \label{fig: new coar start sketch}
\end{figure}

As mentioned in Section \ref{sec: problem statement}, topology optimisation of heat conduction problems often results in structures containing very small features, which will be poorly resolved on the coarse grids if too much $x$-coarsening is applied. Therefore, it might be advantageous to modify the coarsening strategy such that it performs $t$-coarsening when the spatial resolution on the coarse grids becomes too poor. Such a modification is presented in Algorithm \ref{alg: coar strat for small features}. This coarsening strategy is identical to Algorithm \ref{alg: coar strat for high contrast}, except on line \ref{algline: nelmin modification} where the if-statement has been modified to perform an additional check. Specifically, it computes the number of elements which will be on level $l+1$ if it applies $x$-coarsening, which is $N_\mathrm{el}^{(l)}/2$. It then checks if this is less than a user-defined threshold denoted $M$. If the number of elements is going to become less than $M$, then it applies $t$-coarsening. Otherwise, the coarsening direction is chosen based on $\lambda_\mathrm{eff}$ like in Algorithm \ref{alg: coar strat for high contrast}. As such, this coarsening strategy ensures that $N_\mathrm{el} \geq M$ on every level. Figure \ref{fig: new coar start sketch} shows an example of a ``coarsening path" which can be generated by this coarsening strategy.

\begin{algorithm}
\caption{Coarsening strategy for high-contrast problems with small features.}
\label{alg: coar strat for small features}
\begin{algorithmic}[1]
    \STATE Construct level 1 (the finest level)
    \FOR {$l = 1, \hdots, N_l-1$}
        \STATE $D_\mathrm{eff}^{(l)} \gets \sqrt{ \min_e\left(D_e^{(l)}\right) \max_e\left(D_e^{(l)}\right) } $
        \STATE $\lambda_\mathrm{eff}^{(l)} \gets D_\mathrm{eff}^{(l)} \dfrac{\Delta t^{(l)}}{  \left(\Delta x^{(l)}\right)^2  }$
        \IF{$\lambda_\mathrm{eff}^{(l)} < \lambda_\mathrm{crit}$ \textbf{or} $N_\mathrm{el}^{(l)} / 2 < M$} \label{algline: nelmin modification}
            \STATE Construct level $l+1$ by coarsening level $l$ in the $t$-direction
        \ELSE
            \STATE Construct level $l+1$ by coarsening level $l$ in the $x$-direction
        \ENDIF
    \ENDFOR
\end{algorithmic}
\end{algorithm}

\subsection{Tests of strategy ensuring spatial resolution}
\label{sec: new coar strat tests}

During testing, mixed results were found regarding the performance of the new coarsening strategy in Algorithm \ref{alg: coar strat for small features}. In this subsection, this will be illustrated by showing examples where there is a clear benefit to using the new strategy as opposed to the old strategy (Algorithm \ref{alg: coar strat for high contrast}), as well as examples where there is no benefit or the benefit is negligible. 

The new proposed coarsening strategy is tested on problem 8 as defined in Section \ref{sec: different redisc methods tests}, except the parameter $F$ (the fraction of the domain occupied by the insulator) is treated as an independent variable. Specifically, $F$ is of the form $2^{-n/3}$ where $n$ set to every integer between $6$ and $24$. For some of these values of $F$, it is found that $\nu=5$ is not enough smoothing steps to stabilise the STMG method. For this reason, $\nu$ is set to 20 instead in all tests of the new coarsening strategy. The parameter $\lambda_\mathrm{crit}$ is set to 0.25, and both the CK- and CR-rediscretisation methods are applied to this problem. 

The number of levels is set to 6, which means that the number of spatial elements on the coarsest level is going to be $256 / 2^{6-1} = 8$ if $x$-coarsening is applied on every level. As such, if $M < 8$, then this will have the same effect as setting $M=8$, because the coarsening strategy can never reach a level where $N_\mathrm{el} < 8$.  For this reason, the new coarsening strategy is tested for $M=8$, $16$, $32$, $64$, and $128$, and the case where $M=8$ is labelled as ``$M \leq 8$" to indicate that identical results would be obtained if $M < 8$. It is also noted that setting $M \leq 8$ is equivalent to using the old coarsening strategy. 

\begin{figure}
    \centering
    \subfloat[CK-rediscretisation method]{\includegraphics[width=0.75\linewidth]{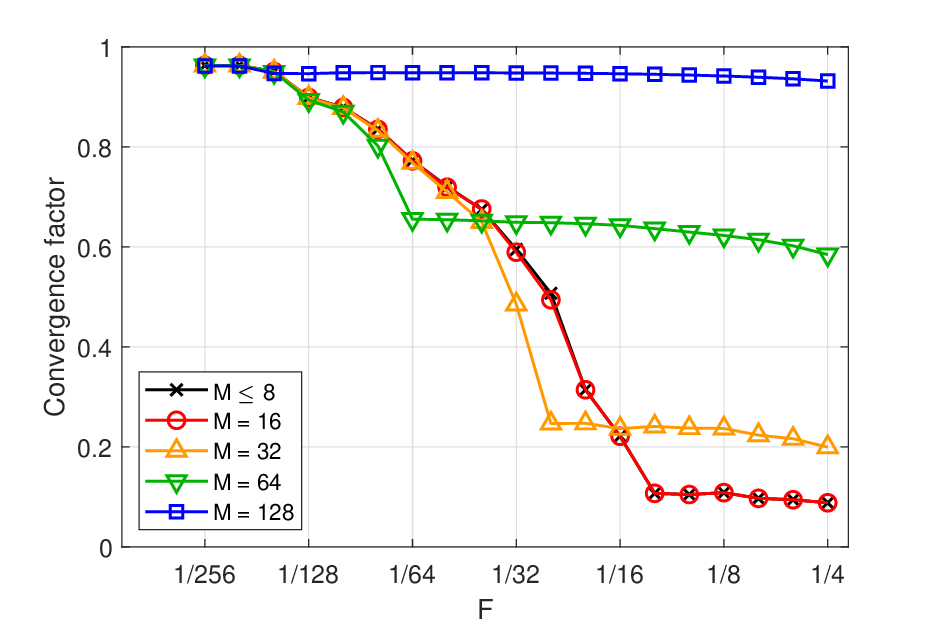}\label{fig: prob 10 CK}}
    
    \subfloat[CR-rediscretisation method]{\includegraphics[width=0.75\linewidth]{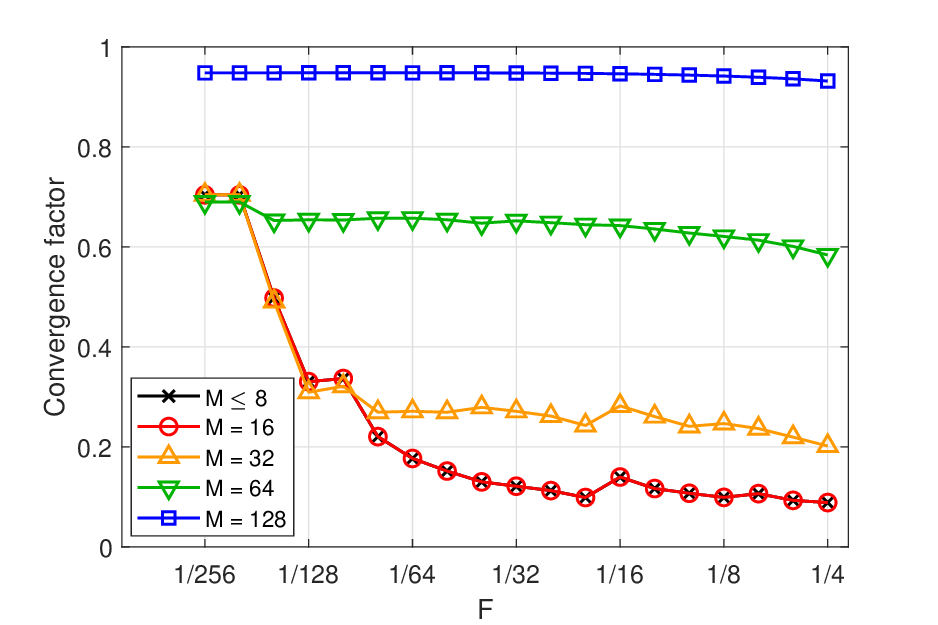}\label{fig: prob 10 CR}}
    \caption{Results of the coarsening strategy defined in Algorithm \ref{alg: coar strat for small features} when using the CK- and CR-rediscretisation methods as defined in Section \ref{sec: different redisc methods defs}. It is applied to a test problem containing a thin insulating structure, and $F$ is the fraction of the domain occupied by this structure. In this context, $M \leq 8$ is equivalent to the previous coarsening strategy in Algorithm \ref{alg: coar strat for high contrast}. }
\end{figure}

The results for the CK-rediscretisation method are shown in Figure \ref{fig: prob 10 CK}. Firstly, it is seen that setting $M \leq 8$ yields almost the same convergence factors as setting $M=16$. Therefore, it can be said that the new strategy is not particularly beneficial if $M=16$. However, there are some cases where there is a significant benefit in setting $M=32$ or $64$. In particular, in the area where $F \approx 1/32$, there is a clear benefit in setting $M=32$, and in the area where $F \approx 1/64$, there is a clear benefit in setting $M=64$. However, this pattern does not extend to the region where $F \approx 1/128$. Instead, the optimal choice of $M$ is 64 at $F=1/128$, while in the region where $F < 1/128$, it is optimal to set $M=128$. However, the benefit in setting $M=128$ is negligible. Based on these observations, it can be said that there is a correlation between the feature size, $F$, and the optimal value of $M$. Specifically, there are many cases where it is optimal to set $M \approx 1/F$. This implies that the element width, $\Delta x$, should be close to the width of the small insulating feature. 

The results for the CR-rediscretisation method are shown in Figure \ref{fig: prob 10 CR}. For this rediscretisation method, the results for $M \leq 8$ and $M = 16$ are exactly equal to each other, because the coarsening strategies chose the same coarsening directions at every level. Apart from that, it can be seen that there are some cases where it is beneficial to set $M=32$ or $64$, but the benefit is less significant than it was for the CK-method. Also, these benefits do not appear at the same values of $F$ as they did in the previous case. For example, around $F \approx 1/256$ it is optimal to set $M=64$, while at $F\approx 1/128$, it is optimal to set $M=32$. Meanwhile, for $F \geq 1/64$, there is no benefit in using the new strategy. Additionally, it can be noted that setting $M=128$ makes the convergence factors significantly worse for all values of $F$. 

\subsection{Discussion of strategy ensuring spatial resolution}

It is seen that the results regarding the new coarsening strategy are mixed, as stated earlier. This can be explained by the fact that there are two competing factors which impact the performance. The first factor is that the multigrid method benefits from being able to resolve small structures on the coarse grids, which means that it is disadvantageous to perform too much $x$-coarsening. However, the second factor is that it is very disadvantageous to apply $t$-coarsening when $\lambda_\mathrm{eff}$ is large, as shown in Figure \ref{fig: prob 1 to 6 results}. Additionally, it is noted that $\lambda_\mathrm{eff}$ will always increase when $t$-coarsening is applied, since $\Delta t$ is doubled while $D_\mathrm{eff}$ and $\Delta x$ are unchanged. Therefore, the new coarsening strategy easily leads to situations where $t$-coarsening is applied on levels where $\lambda_\mathrm{eff}$ is very large, which makes the performance of the new strategy even worse. 

The above factors probably explain the patterns observed in the case with the CK-rediscretisation method, where it was beneficial to set $M \approx 1/F$. It is likely that it benefits from having just enough elements on the coarse levels to resolve the small insulating feature, but no more than that, otherwise the anisotropy becomes the dominant factor. Meanwhile, it is noted that the CR-rediscretisation method is better than the CK-method for the considered problems, because it better represents the relevant physics on the coarse levels, as discussed in Section \ref{sec: redisc methods discussion}. As such, the ability to resolve small structures is probably less relevant, which means that it becomes more important to consider the anisotropy, which likely explains why there is little-to-no benefit in using the new coarsening strategy when the CR-method is applied.

Additionally, it is noted that for very small feature sizes ($F \leq 1 / 128$), the CR rediscretisation method yields convergence factors which are significantly better than those of the CK rediscretisation method. This emphasises the need for appropriate rediscretisation methods when working with problems which include small features.

\section{Space-time multigrid method applied to an optimisation problem}
\label{sec: opt prob application}

To investigate the robustness of the STMG method, it is applied in the context of an optimisation problem which is, in some ways, similar to practical topology optimisation problems.

\subsection{Optimisation problem}

The objective functional, $\Theta$, for the considered optimisation problem is defined to be proportional to the time-integral of the thermal compliance, like so:
\begin{align}
    \Theta = \frac{1}{\Theta_\mathrm{ref}} \int_{0}^{t_T} \int_{0}^{L} q(x, t) T(x, t) \, \mathrm{d}x \, \mathrm{d}t 
\end{align}
where $\Theta_\mathrm{ref} = 10^6 \, \mathrm{JK/m^2}$ is a normalisation constant which makes it so that $\Theta$ is on the order of 1. This type of objective functional has been referred to as the ``transient thermal dissipation efficiency" \cite{Wu2021_TTDE}. In the context of the BE-FE discretisation, the objective is approximated using the following formula:
\begin{align}
    \Theta = \mbf{b}\Tr \mbf{u} \Delta t / \Theta_\mathrm{ref}
\end{align}
The optimisation problem is then formulated as the following:
\begin{subequations} 
\label{eq: opt problem}
\begin{align}
    \min_{\boldsymbol{\chi}} \quad & \Theta = \mbf{b}\Tr \mbf{u} \Delta t / \Theta_\mathrm{ref}
    \\
    \textrm{s.t.} \quad 
    & \mbf{J} \mbf{u} = \mbf{b} \label{eq: opt prob primal prob}
    \\
    & 0 \leq \chi(x) \leq 1 \quad \forall  \, x \in [0,L]  \label{eq: opt prob box constraint}
    \\
    & \int_0^L \chi(x) \, \mathrm{d}x \leq L/2 \label{eq: opt prob volume constraint}
\end{align}
\end{subequations}
where $\boldsymbol{\chi}$ is the vector containing the discretised degrees of freedom of the design field, $\chi(x)$. For this optimisation problem, the parameter values chosen for problem 7 in Section \ref{sec: different redisc methods tests} are adopted, meaning that $L = 0.1\,\mathrm{m}$, $t_T = 10\,\mathrm{s}$, $N_\mathrm{el}=N_t=256$, the heat load is given by Equation (\ref{eq: heat load def}), and the material parameters are listed in Table \ref{tab: prob 7-9 thermal properties}. This optimisation problem is similar to practical topology optimisation problems in the following ways:
\begin{itemize}
    \item It is subject to a box constraint on the design field (Equation (\ref{eq: opt prob box constraint})) and a volume constraint (Equation (\ref{eq: opt prob volume constraint}));
    \item The material parameters are penalised using a SIMP scheme, as defined in Section \ref{sec: simp scheme};
    \item For the specific parameters chosen for this problem, the optimiser is observed to converge towards a design field which contains many small features and is almost entirely 0-1, meaning that $\chi$ is either 0 or 1 for most of the elements. This will be shown in Section \ref{sec: opt prob results}.  
\end{itemize}
However, it is also dissimilar to practical topology optimisation problems in the following ways:
\begin{itemize}
    \item The problem only considers one spatial dimension;
    \item No filter is applied to the design, despite the fact that filters are often necessary for regularising the results when performing topology optimisation \cite{bourdin2001_topOptFilter}; 
    \item If $t_T$ is increased (to $100 \, \mathrm{s}$, for example), then the optimiser converges to a design field containing many greyscale elements, meaning elements where $0 < \chi < 1$. As such, it can be said that the penalisation scheme used here does not work reliably for this type of optimisation problem when considering only one spatial dimension. 
\end{itemize}
Despite these differences, this optimisation problem is taken to be sufficiently similar to practical topology optimisation problems.

\subsection{Optimisation algorithm}
\label{sec: opt prob algorithm}

The optimisation problem in Equation (\ref{eq: opt problem}) is solved using a nested approach where, in each optimisation cycle, the physics is solved using the STMG method devised in this paper, after which the design field is updated using the Method of Moving Asymptotes \cite{MMA_original}. To obtain the sensitivities of the objective with respect to the design field, the discrete adjoint method is used \cite{Michaleris1994_discAdjForTransient}, wherein the adjoint solution vector, $\boldsymbol{\Lambda}$, is defined by the following equation: 
\begin{align}
    \mbf{J}\Tr \boldsymbol{\Lambda} &= \left( \frac{\partial\Theta}{\partial \mbf{u}} \right)\Tr = \mbf{b} \Delta t / \Theta_\mathrm{ref}
\end{align}
This equation is also solved using the STMG method. In this context, the system matrix on $l$\tss{th} level of the STMG method is set to $(\mbf{J}^{(l)})\Tr$. These transposed system matrices can be interpreted as representing modified physics where the direction of time is reversed. As such, if causal interpolation is used, then it makes sense to define the prolongation operator such that it transfers information backwards in time, not forwards, when the adjoint problem is solved. However, in practice it is observed that the STMG method solving the adjoint problem performs reasonably well, even when using the prolongation operator which sends information only forwards in time. This will be shown in Section \ref{sec: opt prob results}. For this reason, the prolongation operators for the primal and adjoint problems are set equal to each other in every test. When solving the adjoint problem, the relative residual is defined as the following: 
\begin{align}
    r_n &= \frac{\lVert \mbf{J}\Tr \boldsymbol{\Lambda}_n - \mbf{b} \Delta t / \Theta_\mathrm{ref} \rVert}{\lVert \mbf{b} \Delta t / \Theta_\mathrm{ref} \rVert} 
\end{align}
where $\boldsymbol{\Lambda}_n$ is $\boldsymbol{\Lambda}$ after the $n$\tss{th} cycle. The termination criterion in Equation (\ref{eq: stmg termination criterion}) is also used for the adjoint problem. 

The used coarsening strategy is a variation of Algorithm \ref{alg: coar strat for high contrast} which has been changed such that the prolongation and restriction operators only have to be defined once during the optimisation process. It is identical to Algorithm \ref{alg: coar strat for high contrast}, except the effective diffusivity on every level is defined as: 
\begin{align}
    D_\mathrm{eff} &\equiv \sqrt{ \min_{\chi \in [0, 1]}D_\mathrm{SIMP}(\chi) \max_{\chi \in [0, 1]}D_\mathrm{SIMP}(\chi) }
\end{align}
where the function $D_\mathrm{SIMP}(\chi)$ is defined as:
\begin{align}
    D_\mathrm{SIMP}(\chi) \equiv k_\mathrm{SIMP}(\chi) / c_\mathrm{SIMP}(\chi)
\end{align}
This definition of $D_\mathrm{eff}$ is independent of the considered design field, since it only depends on the properties of the two materials and the penalisation scheme. Therefore, the coarsening strategy will always choose the same coarsening path at every optimisation cycle, meaning that the coarsening path, prolongation operators, and restriction operators only have to be defined once at the start of the optimisation process. This has the added benefit that it is only necessary to allocate memory for the system matrices and solution vectors once at the start of the optimisation process (however, these still have to be assembled and solved for at each optimisation cycle, since the design field changes). Additionally, it removes the ambiguity in how to define $\lambda_\mathrm{eff}$ on the coarse grids when Galerkin projection is used. It is easy to evaluate $D_\mathrm{eff}$ for the specific optimisation problem considered here, because the extrema of $D_\mathrm{SIMP}(\chi)$ are the following:
\begin{subequations}
\label{eq: D_SIMP extrema}
\begin{align}
    \min_{\chi \in [0, 1]}D_\mathrm{SIMP}(\chi) &= D_\mathrm{SIMP}(0) = k_\mathrm{ins} / c_\mathrm{ins} \\
    \max_{\chi \in [0, 1]}D_\mathrm{SIMP}(\chi) &= D_\mathrm{SIMP}(1) = k_\mathrm{con} / c_\mathrm{con} 
\end{align}
\end{subequations}
Note that Equation (\ref{eq: D_SIMP extrema}) does not hold for all material parameters and penalisation schemes, because there may be local maxima or minima of the function $D_\mathrm{SIMP}(\chi)$ in the interval $0 < \chi < 1$. 

As in Section \ref{sec: new coar strat tests}, the number of levels is set to $N_l=6$, the parameter $\lambda_\mathrm{crit}$ is set to $0.25$, and $\nu$ is set to $20$. Three different rediscretisation methods are tested in this context: the BR-, CR-, and CP-method. In the context of optimisation problems, iterative methods for solving the physics (like multigrid methods) often benefit from using ``warm restarts" \cite{Minion2018_pfasstOptCtrl, appel2024_oneShotPar}, where the guess for the solution at each optimisation cycle is the solution found at the previous cycle. For the sake of investigating the robustness of the considered STMG method, it will be tested using both warm restarts and ``cold restarts", where $\mbf{u} = \mbf{0}$ and $\boldsymbol{\Lambda} = \mbf{0}$ are the guesses for the solutions at each optimisation cycle. 

At the start of the optimisation process, the design field is initialised as being $\chi_e = 1/2$ at every element. The optimisation process is terminated when the relative change in $\Theta$ has been less than $0.1\%$ for 5 consecutive optimisation cycles. 


\subsection{Optimisation results}
\label{sec: opt prob results}

\begin{figure}
    \centering
    \includegraphics[width=0.8\linewidth]{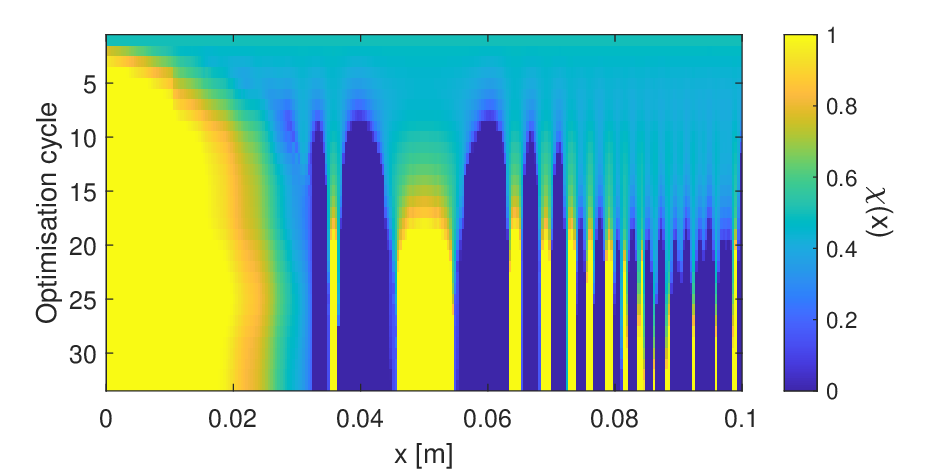}
    \caption{Evolution of the design field over the optimisation cycles of the algorithm described in Section \ref{sec: opt prob algorithm}. }
    \label{fig: optprob desfield evol}
\end{figure}

The evolution of the design was observed to be independent of the rediscretisation method and the use of warm or cold restarts. Figure \ref{fig: optprob desfield evol} shows the evolution of the design field over the optimisation cycles. It is seen that the design is uniform at the start, after which the contrast increases and it gains finer details. As stated earlier, the design field at the final iteration contains many small features and is mostly 0-1. The main exception to this is in the interval $0.015\,\mathrm{m} < x < 0.035\,\mathrm{m}$ where there is a smooth transition in the design field.

\begin{figure}
    \centering
    \subfloat[BR rediscretisation method.]{\includegraphics[width=0.7\linewidth]{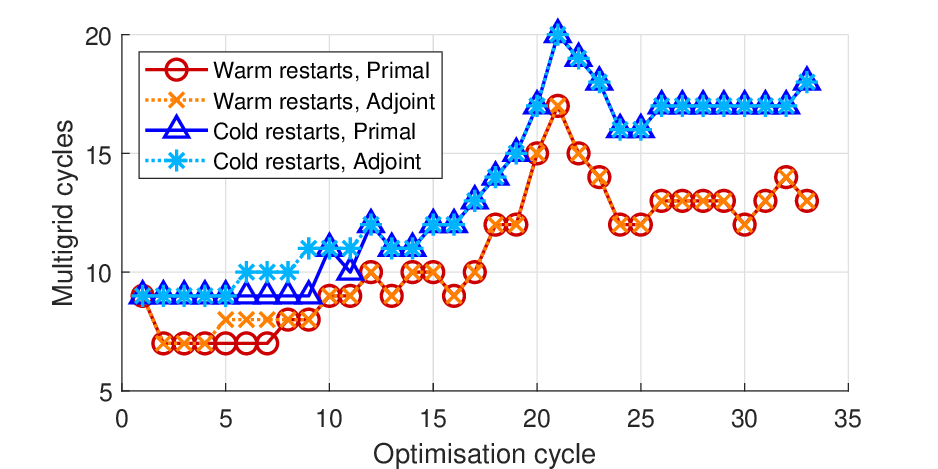}}
    
    \subfloat[CR rediscretisation method.]{\includegraphics[width=0.7\linewidth]{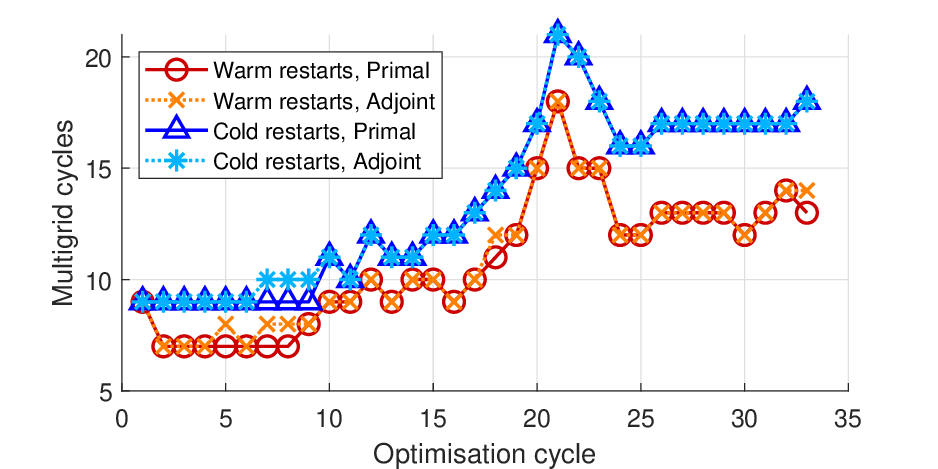}}
    
    \subfloat[CP rediscretisation method.]{\includegraphics[width=0.7\linewidth]{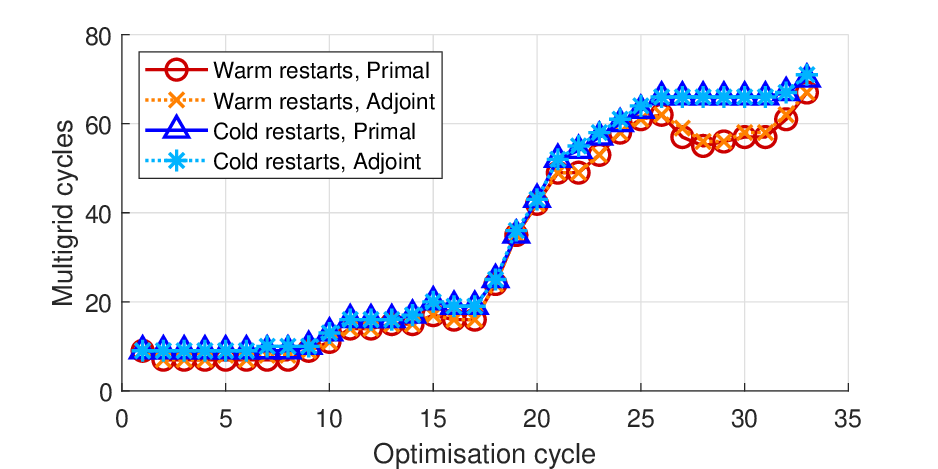}}
    \caption{Number of multigrid cycles per optimisation cycle of the optimisation algorithm described in Section \ref{sec: opt prob algorithm}, shown for both types of restarts, for both the primal and adjoint problems, and for three different rediscretisation methods. }
    \label{fig: optprob MG cycles per opt cycle}
\end{figure}

Figure \ref{fig: optprob MG cycles per opt cycle} shows the number of STMG cycles at each optimisation cycle for both the primal and adjoint problems for each of the rediscretisation methods. In each case, the STMG converges in fewer than 80 cycles at every point in the evolution of the design. As such, the STMG method is deemed to be sufficiently robust, since it is able to converge for a variety of designs with different material distributions, different amounts of contrast, and different gradients in the material parameters. It is also seen that the number of STMG cycles generally increases during the optimisation process. This is to be expected, since the design evolves to have higher contrast and finer details. Additionally, the number of cycles is reduced when using warm restarts instead of cold restarts, as expected. 

It is also seen that the number of STMG cycles for the adjoint problem is very close to the number of cycles for the primal problem in every case. This shows that it is not necessary for the direction of time for the causal prolongation operator to be the same as the direction of time for the physics being solved. This is especially noteworthy in the context of the CP-method, because the prolongation operator is directly involved in the construction of the system matrices on the coarse levels in this case. 

Despite being very close to each other, it is seen that the total number of cycles for the adjoint problem is consistently slightly larger than the total number of cycles for the primal problem. This can not be explained by a mismatch in the direction of time between the adjoint physics and the prolongation operator, since this is also observed for bilinear space-time interpolation, for which there is no intrinsic direction of time. It is also not due to differences in the source terms between the primal and adjoint problems, because the source term for the adjoint problem is a scaled version of the source term for the primal problem. The considered STMG method is linear, since the smoothers and coarse grid corrections are linear operations, so the convergence rate is independent of the scaling of the source term. The convergence criterion is also independent of the scaling of the source term, since the convergence criterion is based on a relative residual. Therefore, the greater number of cycles for the adjoint problem must be attributed to the difference in the system matrices between the primal and adjoint problems, since this is the only other factor that is different between the two problems.

\section{Conclusion}
\label{sec: conclusion}

This paper has aimed to formulate Space-Time MultiGrid (STMG) methods which are suitable for performing density-based topology optimisation of transient heat conduction problems. This was done by using numerical experiments to investigate the performance of various coarsening strategies, interpolation methods, and matrix reassembly methods when applied to problems with small features and high contrast in the material properties. This was done under the constraint of using only uniform Cartesian space-time meshes and geometric multigrid methods. 

In cases with high contrast in the material properties, it becomes ambiguous whether it is optimal to apply semi-coarsening in space ($x$-coarsening) or in time ($t$-coarsening), because the anisotropy parameter becomes variable over space. As a solution to this problem, this paper proposed a coarsening strategy where the coarsening direction is decided by evaluating an effective anisotropy parameter on each level. Through trial and error, it was found that it is beneficial to define this effective anisotropy parameter as being proportional to the geometric mean of the minimum and maximum thermal diffusivity. This was demonstrated by applying the two-grid STMG method to six different test problems with varying material distributions and varying contrast in the thermal conductivity and heat capacity. These experiments showed that the effective anisotropy parameter is a reasonably reliable indicator for whether it is optimal to use $x$-coarsening or $t$-coarsening. However, the authors do not know if the presented definition of the effective anisotropy parameter will work reliably for other choices of discretisations and multigrid method components. 

This paper considered two different methods of interpolating the temperature field and four different methods of reassembling the system matrices on the coarse levels of the STMG method. Each of the eight combinations of the interpolation and reassembly methods were tested on three different test problems. It was found that if bilinear space-time interpolation was combined with Galerkin projection of the system matrices, then the considered STMG method became highly unstable when there are many levels. Aside from that, the other seven combinations were stable. It was found that for a problem without small features, the interpolation method for the temperature became the dominant deciding factor for the performance of the STMG method. Specifically, it was better to use causal interpolation, which does not transfer information backwards in time, instead of bilinear space-time interpolation. However, for problems including small features, it was found that the matrix reassembly method was the dominant deciding factor for the performance. For this, the best reassembly method was found to be one where the thermal resistivities on the coarse levels were computed by averaging the resistivity on the finer levels. However, this is likely a consequence of the fact that only one dimension of space was considered in the test problems. For practical topology optimisation problems in two or three dimensions of space, it will likely be better to use averaging of the conductivity or Galerkin projection. 

This paper also considered another coarsening strategy which was intended to work well for problems containing small features. This coarsening strategy was formulated such that it applies $t$-coarsening instead of $x$-coarsening if the spatial resolution becomes too poor on the coarse levels. Mixed results were found for this alternative coarsening strategy. In some cases, there was a clear benefit in using the alternative strategy over the previous strategy, while in other cases there was little or no benefit. This was explained by the fact that there are two competing factors which affect the performance. The first factor is the ability to resolve small structures on the coarse grids. The second factor is that it becomes disadvantageous to use $t$-coarsening if the effective anisotropy parameter becomes too large. 

To investigate the robustness of the devised STMG method, it was applied to an optimisation problem which was similar to a topology optimisation problem, in the sense that the optimiser converged to a design with high contrast and small features. However, it was also dissimilar to practical topology optimisation problems regarding heat conduction, since it only considered one dimension of space. The STMG method was deemed to be sufficiently robust, because it converged for every design that was generated during the optimisation process. Also, it was beneficial to use warm restarts for the STMG method as opposed to cold restarts, as expected. It was also found that the STMG method worked well when applied to the adjoint problem associated with the optimisation problem. It even worked well when the chosen prolongation operator only sends information forwards in time, despite the fact that the direction of time for the adjoint problem is backwards. This was also observed when using Galerkin projection, which is especially noteworthy, since the prolongation operator is directly involved in the construction of the system matrices on the coarse levels in this case.

\section*{CRediT authorship contribution statement}

\textbf{Magnus Appel:} Conceptualisation, Methodology, Software, Investigation, Formal analysis, Writing - Original Draft, Visualisation. 
\textbf{Joe Alexandersen:} Conceptualisation, Methodology, Software, Writing - Review \& Editing, Supervision, Funding Acquisition. 

\section*{Declaration of competing interests}

The authors declare that they have no known competing financial interests or personal relationships that could have appeared to influence the work reported in this paper.

\section*{Acknowledgements}

The work has been funded by Independent Research Fund Denmark (DFF) through a Sapere Aude Research Leader grant (3123-00020B) for the COMFORT project (COmputational Morphogenesis FOR Time-dependent problems). DFF had no involvement in this work other than funding. 

The implementation of the space-time multigrid methods described in this paper are modifications of the multigrid method MATLAB code published by \citet{amir2014_topOptMG}. We thank them for making the code available.

\appendix

\section{Handling of Dirichlet boundary conditions}
\label{app: handling diri bc}

To enforce Dirichlet boundary conditions, a diagonal matrix, $\mbf{B}$, is introduced, the entries of which are defined like so: 
\begin{align}
    B_{i,j} &= 
    \begin{cases}
        0 & \text{if $i \neq j$ or the $i$\tss{th} degree of freedom is at a Dirichlet boundary condition}  \\ 
        1 & \text{otherwise}
    \end{cases}
\end{align}
In this context, the initial condition is also considered a Dirichlet boundary condition. The system matrix and source term are then defined as follows:
\begin{align}
    \label{eq: bc correction for J}
    \mbf{J} &= \mbf{B} \, \mbf{J}_\mathrm{Neu} \, \mbf{B} + W_\mathrm{Diri} \cdot (\mbf{I} - \mbf{B})
    \\
    \mbf{b} &= \mbf{B} \, \mbf{b}_\mathrm{Neu}
\end{align}
where $W_\mathrm{Diri}$ is a scalar, and $\mbf{J}_\mathrm{Neu}$ and $\mbf{b}_\mathrm{Neu}$ are the system matrix and source term, respectively, having been assembled assuming only Neumann boundary conditions. The operation in Equation (\ref{eq: bc correction for J}) eliminates the rows and columns of $\mbf{J}$ associated with the fixed degrees of freedom, after which it inserts the number $W_\mathrm{Diri}$ into the diagonal elements associated with the fixed degrees of freedom. If a direct solver is used to solve the linear system, then the value of $W_\mathrm{Diri}$ is irrelevant, as long as $W_\mathrm{Diri} \neq 0$. However, when using the space-time multigrid method described in this paper, the performance depends on $W_\mathrm{Diri}$, because the restriction operator will mix the residual of the interior points with the residual at the Dirichlet boundary conditions. 

During testing, it was observed that the STMG method becomes unstable for certain values of $W_\mathrm{Diri}$. In some cases, it becomes unstable when $W_\mathrm{Diri}$ is close to 0, while in other cases, it becomes unstable when $W_\mathrm{Diri}$ is slightly negative. To avoid these instabilities, $W_\mathrm{Diri}$ is always defined as: 
\begin{align}
    W_\mathrm{Diri}^{(l)} &= \max_e\left(c_e^{(l)}\right) \Delta x^{(l)} / \Delta t^{(l)} + \max_e\left(k_e^{(l)}\right) / \Delta x^{(l)}
\end{align}
where the superscript $^{(l)}$ indicates that these variables pertain to the $l$\tss{th} level of the multigrid method. The above expression ensures that $W_\mathrm{Diri}$ will always be on the same order of magnitude as the largest coefficients of $\mbf{J}$. 


\bibliographystyle{elsarticle-num-names} 
\bibliography{bib}


\end{document}